\documentclass[conference]{IEEEtran}

\usepackage{todonotes}
\usepackage[hyphens]{url}
\usepackage{paralist}
\usepackage{enumitem}
\usepackage{pifont}

\usepackage{multicol}
\usepackage{balance}

\usepackage{varwidth}
\usepackage{algorithm}
\usepackage[noend]{algpseudocode}

\usepackage{xcolor, listings}
\usepackage{caption}

\usepackage{tikz}
\usepackage{multirow}
\usetikzlibrary{calc,positioning,arrows}

\usepackage[]{pgfplots}
\usepackage{pgfplotstable}
\usepackage{amsthm, amssymb}
\usepackage{amsmath}
\usepackage{subcaption}
\usepackage{wrapfig}

\usepackage{comment}

\usepackage{booktabs}

\newtheorem{example}{Example}
\newtheorem{definition}{Definition}
\newtheorem{theorem}{Theorem}
\newtheorem{lemma}{Lemma}

\usepackage{mathtools}
\usepackage{xspace}

\usepackage{graphicx}
\usepackage{colortbl}
\usepackage{enumitem}

\usepackage{marginnote}

\newcommand{\cas}{\textup{\textsc{CAS}}}

\newcommand{\ourtool}{\textsc{MTC}\xspace}

\newcommand{\sertool}{\textsc{MTC-SER}\xspace}
\newcommand{\sitool}{\textsc{MTC-SI}\xspace}
\newcommand{\ssertool}{\textsc{MTC-SSER}\xspace}

\newcommand{\co}{\red{co}}
\newcommand{\wrrelation}{\red{wr}}
\newcommand{\sorelation}{\red{so}}

\newcommand{\sccond}{\textup{\textsc{SC}}}
\newcommand{\lin}{\textup{\textsc{LIN}}}
\newcommand{\iso}[1]{\textsc{#1}}
\newcommand{\sseriso}{\textsc{Strict Serializability}\xspace}
\newcommand{\seriso}{\textsc{Serializability}\xspace}
\newcommand{\siiso}{\textsc{Snapshot Isolation}\xspace}

\newcommand{\sser}{\textup{\textsc{SSER}}\xspace}
\newcommand{\ser}{\textup{\textsc{SER}}\xspace}
\newcommand{\si}{\textup{\textsc{SI}}\xspace}

\newcommand{\nphard}{\textsf{\textup{NP-hard}}}
\newcommand{\nphardness}{\textsf{\textup{NP-hardness}}}
\newcommand{\npc}{\textsf{\textup{NP-complete}}}

\newcommand{\red}[1]{\textcolor{red}{#1}}

\newcommand{\emptyseq}{\langle\rangle}

\newcommand{\incell}[2]{\begin{tabular}{@{}#1@{}}#2\end{tabular}}

\newcommand{\np}{\textsf{NP}}

\newcommand{\true}{{\sf true}}
\newcommand{\false}{{\sf false}}
\newcommand{\Key}{{\sf X}} 
\newcommand{\Val}{{\sf V}} 

\newcommand{\h}{\mathcal{H}}

\newcommand{\opset}{\mathit{O}}

\newcommand{\Op}{{\sf Op}}
\newcommand{\readevent}{{\sf R}}
\newcommand{\writeevent}{{\sf W}}
\newcommand{\rwevent}{{\sf R\&W}}
\newcommand{\WriteTx}{{\sf WriteTx}}

\newcommand{\exception}{\textup{\textsc{Divergence}}}

\newcommand{\rel}[1]{\xrightarrow{#1}}
\newcommand{\comp}{\;;\;}

\newcommand{\po}{{\sf po}}

\newcommand{\intaxiom}{\textup{\textsc{Int}}}
\newcommand{\prefixaxiom}{\textup{\textsc{Prefix}}\xspace}
\newcommand{\conflictaxiom}{\textup{\textsc{Conflict}}\xspace}

\newcommand{\cycle}{\mathcal{C}}

\newcommand{\T}{\mathcal{T}}
\providecommand{\G}{}
\renewcommand{\G}{\mathcal{G}}
\renewcommand{\H}{\mathcal{H}}

\newcommand{\SO}{\textup{\textsf{SO}}}
\newcommand{\WR}{\textup{\textsf{WR}}}
\newcommand{\WW}{\textup{\textsf{WW}}}
\newcommand{\RW}{\textup{\textsf{RW}}}
\newcommand{\RT}{\textup{\textsf{RT}}}

\newcommand{\edges}{E}

\newcommand{\set}[1]{\{#1\}}

\newcommand{\hStatex}{\vspace{1pt}}

\newcommand{\valvar}{\mathit{v}}

\renewcommand{\ae}{\mathcal{A}}

\newcommand{\starttime}{\mathit{s}}
\newcommand{\committime}{\mathit{f}}
\newcommand{\earlycommittime}{\mathit{min\_f}}
\newcommand{\nextinchain}{t}

\newcommand{\checksser}{\textup{\textsc{CheckSSER}}}
\newcommand{\checkser}{\textup{\textsc{CheckSER}}}
\newcommand{\checksi}{\textup{\textsc{CheckSI}}}
\newcommand{\consdep}{\textsc{BuildDependency}}
\newcommand{\acyclic}{\textsc{acyclic}}

\newcommand{\circled}[1]{{\textcircled{\small #1}}}

\newcommand{\chain}{\mathit{chain}}
\newcommand{\vlcas}{\textup{\textsc{VL-LWT}}}

\newcommand{\rt}{\mathit{rt}}

\newcommand{\optstyle}[1]{\boxed{\colorbox{green!30}{$#1$}}}

\newcommand{\thinairread}{\textsc{ThinAirRead}\xspace}
\newcommand{\abortedread}{\textsc{AbortedRead}\xspace}
\newcommand{\futureread}{\textsc{FutureRead}\xspace}
\newcommand{\notmyownwrite}{\textsc{NotMyOwnWrite}\xspace}
\newcommand{\notmylastwrite}{\textsc{NotMyLastWrite}\xspace}

\newcommand{\intermediateread}{\textsc{IntermediateRead}\xspace}
\newcommand{\nonrepeatableread}{\textsc{NonRepeatableReads}\xspace}
\newcommand{\nonmonoread}{\textsc{NonMonotonicRead}\xspace}
\newcommand{\sessionguaranteeviolation}{\textsc{SessionGuaranteeViolation}\xspace}
\newcommand{\causalityviolation}{\textsc{CausalityViolation}\xspace}
\newcommand{\lostupdate}{\textsc{LostUpdate}\xspace}
\newcommand{\longfork}{\textsc{LongFork}\xspace}
\newcommand{\writeskew}{\textsc{WriteSkew}\xspace}
\newcommand{\fracturedread}{\textsc{FracturedRead}\xspace}

\newcommand{\reviewright}[4]{#4}
\newcommand{\reviewleft}[4]{#4}

\begin{document}

\title{Boosting End-to-End Database Isolation Checking
  via Mini-Transactions (Extended Version)}

\author{\IEEEauthorblockN{Hengfeng Wei, Jiang Xiao, Na Yang}
\IEEEauthorblockA{\textit{State Key Laboratory for Novel Software Technology}\\\textit{Nanjing University} \\
hfwei@nju.edu.cn \\ \{mg21330063, nayang\}@smail.nju.edu.cn}
\and
\IEEEauthorblockN{Si Liu, Zijing Yin}
\IEEEauthorblockA{\textit{ETH Zurich} \\
\{si.liu, zijing.yin\}@inf.ethz.ch\\
}
\and
\IEEEauthorblockN{Yuxing Chen, Anqun Pan}
\IEEEauthorblockA{\textit{Tencent Inc.} \\
\{axingguchen, aaronpan\}@tencent.com}
}

\maketitle
\thispagestyle{plain}
\pagestyle{plain}


\begin{abstract}
Transactional isolation guarantees are crucial for database correctness.
However, recent studies have uncovered numerous isolation bugs
in production databases.
The common black-box approach to isolation checking
stresses databases with large, concurrent, randomized transaction workloads
and verifies whether the resulting execution histories
satisfy specified isolation levels.
For strong isolation levels such as strict serializability,
serializability, and snapshot isolation,
this approach often incurs significant end-to-end checking overhead
during both history generation and verification.

We address these inefficiencies through the novel design of Mini-Transactions (MTs).
MTs are compact, short transactions
that execute much faster than general workloads,
reducing overhead during history generation
by minimizing database blocking and transaction retries.
By leveraging MTs' read-modify-write pattern,
we develop highly efficient algorithms
to verify strong isolation levels in linear or quadratic time.
Despite their simplicity, MTs are semantically rich
and effectively capture common isolation anomalies described in the literature.

We implement our verification algorithms
and an MT workload generator in a tool called MTC.
Experimental results show that
MTC outperforms state-of-the-art tools
in both history generation and verification.
Moreover, MTC can detect bugs across various isolation levels
in production databases
while maintaining the effectiveness of randomized testing
with general workloads,
making it a cost-effective solution for black-box isolation checking.
\end{abstract}



\section{Introduction}

Databases serve as the backbone of modern web applications.
Transactional isolation levels
(the ``I'' in ACID -- Atomicity, Consistency, Isolation, and Durability~\cite{TIS:Book2001}),
such as \sseriso (\sser), \seriso (\ser), and \siiso (\si)~\cite{CritiqueANSI:SIGMOD1995},
are crucial for ensuring database correctness.
However, recent studies~\cite{Elle:VLDB2020,PolySI:VLDB2023,Complexity:OOPSLA2019,plume,txcheck}
have uncovered numerous isolation bugs in many production databases,
raising concerns about whether these databases
actually uphold their promised isolation guarantees in practice.
Isolation bugs are notoriously difficult to detect,
as they often produce incorrect results without any explicit errors.
For example, an \ser violation in PostgreSQL remained undetected
for nearly a decade before being identified recently~\cite{jepsen-pg}.

In recent years, a range of sophisticated
checkers~\cite{Elle:VLDB2020,PolySI:VLDB2023,Complexity:OOPSLA2019,plume,Cobra:OSDI2020,Viper:EuroSys2023,porcupine}
has emerged to verify database isolation guarantees.
These checkers operate in a \emph{black-box} fashion,
as database internals are often unavailable or opaque.
To extensively stress test databases,
they rely on randomized workload generators
that produce large, concurrent transaction sets,
aiming to either increase the likelihood of exposing isolation bugs
or to build confidence in their absence.
Each generated workload is \emph{general},
typically involving dozens of clients,
tens of thousands of transactions per client,
and dozens of operations per transaction.
Subsequently, these checkers collect database execution histories
to verify compliance with the specified isolation level.
The Cobra~\cite{Cobra:OSDI2020} and PolySI~\cite{PolySI:VLDB2023} checkers
represent these histories as transactional dependency graphs~\cite{Adya:PhDThesis1999,AnalysingSI:JACM2018}
and leverage off-the-shelf solvers
to identify specific cycles that indicate isolation violations.

\paragraph*{\textbf{\textit{Challenges}}}
The randomized testing approach using general transaction (GT) workloads
frequently encounters efficiency challenges in practice,
primarily due to two factors.

First, isolating a large number of highly concurrent GTs
imposes significant \emph{execution overhead} on databases.
This overhead is particularly exacerbated under strong isolation levels,
including \sser, \ser, and \si~\cite{HAT:VLDB2013,noc-noc}.
For example, with optimistic concurrency control~\cite{Bernstein:Book1987},
databases may abort more transactions due to conflicts,
such as concurrent transactions attempting to write to the same object.
To obtain a history with sufficiently many committed transactions,
the system may require time-intensive retries.
Alternatively, under pessimistic concurrency control~\cite{Bernstein:Book1987},
acquiring and releasing locks to resolve conflicts
among concurrent transactions also incurs significant time costs.

Second, large histories composed of GTs
often result in extensive, dense transactional dependency graphs~\cite{Adya:PhDThesis1999,AnalysingSI:JACM2018, Complexity:OOPSLA2019}.
While such graphs are effective for detecting isolation bugs,
they also impose substantial \emph{verification overhead} on isolation checkers,
which must handle the complexity of encoding and solving dependency constraints
or traversing these graphs to identify cycles.
This verification overhead is further amplified by
the inherently high complexity of verifying strong isolation levels
like SSER, SER, and \si,
which have been shown to be NP-hard in general~\cite{Complexity:OOPSLA2019,SER:JACM1979}.
Notably, a checker's efficiency can greatly
impact its effectiveness in bug detection,
given the often limited resources
such as time and memory~\cite{PolySI:VLDB2023,plume, Cobra:OSDI2020}.

A straightforward approach to addressing the above challenges
is to randomly generate smaller transaction workloads,
e.g., with each transaction comprising only a few operations.
However, this presents a dilemma.
On one hand, such workloads could mitigate inefficiency,
as shorter transactions generally execute faster
and have lower abort rates.
On the other hand, these short transactions may not be
as effective as large GTs in uncovering isolation bugs.
In fact, randomly generated, arbitrarily short transactions
may even fail to capture certain data anomalies.
For example, the \writeskew anomaly, which violates \ser,
requires both concurrent transactions
to read and subsequently write to both objects $x$ and $y$,
as illustrated in Figure~\ref{fig:write-skew}
(see Section~\ref{ss:capturing-anomalies}).

\paragraph*{\textbf{\textit{Our Approach}}}
In this paper, we propose a unified solution to this dilemma
through the novel design of \emph{Mini-Transactions} (MTs),
which addresses both types of inefficiencies
while preserving the effectiveness of randomized testing
in detecting isolation bugs.

An MT is a compact transaction,
containing no more than four read/write operations
(see Section~\ref{sec:mini}),
which we employ in both the \emph{history generation}
and \emph{history verification} stages.
As MTs are short, they can be executed much faster than GTs
within a database,
leading to more efficient processing and reduced overhead.
Specifically, shorter transactions minimize database blocking
and exhibit lower abort rates.

However, even with these improvements in execution time,
efficient verification against strong isolation levels remains challenging
due to the potential for large transaction numbers,
which generate dense dependency graphs.
To address this issue, we design MTs to adhere to
the \emph{read-modify-write} (RMW) pattern,
where each write operation is preceded by a read operation
on the same object.
Furthermore, by ensuring that each MT writes unique values,
a common practice among existing isolation checkers~\cite{Complexity:OOPSLA2019,
Cobra:OSDI2020, PolySI:VLDB2023,Viper:EuroSys2023,Elle:VLDB2020,txcheck},
we enable highly efficient algorithms
in linear or quadratic time (relative to the number of transactions)
to verify strong isolation levels, including SSER, SER, and SI,
over MT histories (see Section~\ref{section:algs}).
Particularly, for MT histories of
\emph{lightweight transactions} (i.e., Compare-And-Set operations),
we develop a linearizability~\cite{Lin:TOPLAS1990} verification algorithm
that operates in linear time.
This design significantly reduces verification overhead.
Notably, we also establish the \nphardness{} of verifying strong isolation levels
for MT histories \emph{without} unique values.

To address the challenge on effectiveness in bug detection,
we demonstrate that, despite their simplicity,
MTs are semantically rich enough to capture
common isolation anomalies that can occur in GT histories.
These encompass all 14 well-documented isolation anomalies
specified in contemporary specification frameworks
for transactional isolation levels~\cite{Adya:PhDThesis1999,
  Framework:CONCUR2015, AnalysingSI:JACM2018, Complexity:OOPSLA2019, plume}.
Such anomalies include \thinairread, \abortedread, \futureread,
\notmylastwrite, \notmyownwrite, \intermediateread, \nonrepeatableread,
\sessionguaranteeviolation, \nonmonoread, \fracturedread,
\causalityviolation, \longfork, \lostupdate, and \writeskew;
see Figure~\ref{fig:anomalies} in Section~\ref{ss:capturing-anomalies}
for illustrations.
We believe this list is exhaustive with respect to
the aforementioned specification frameworks.

\begin{figure}
	\centering
	\includegraphics[scale=0.52]{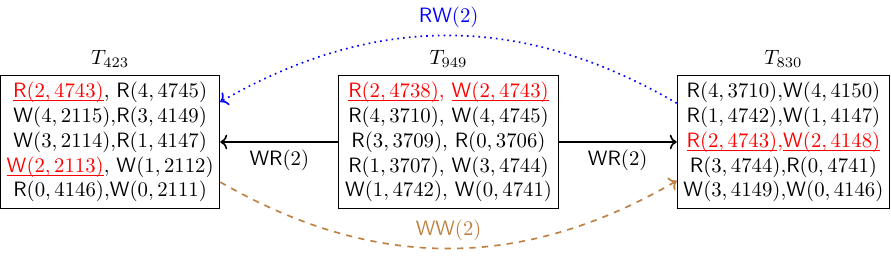}
	\caption{An SI violation (\lostupdate) detected in MariaDB Galera (v10.7.3).
	  The core operations involved are underlined.
		This core comprises only three transactions
		with two operations each, which can be represented by our MTs.}
	\label{fig:mariadb-bug}
\end{figure}

Our insight behind the MT-based characterization of anomalies is that,
while isolation violations have been detected in GT workloads
of substantial size,
these violations are typically due to a small subset of operations
within a limited number of transactions.
For example, Figure~\ref{fig:mariadb-bug} illustrates an SI violation
identified in MariaDB Galera (v10.7.3) using PolySI~\cite{PolySI:VLDB2023},
depicted as a cyclic dependency graph
(see Section~\ref{ss:history-depgraph}).
The corresponding GT workload comprises 10 sessions,
each containing 100 transactions,
with each transaction performing 10 operations.
However, as highlighted in red,
the core of this bug involves only three transactions,
each with two operations,
which can be represented by our MTs.
Specifically, both committed transactions $T_{423}$ and $T_{830}$
read the value written by $T_{949}$ on object $2$
and subsequently write different values.
This leads to a \lostupdate anomaly that violates SI.


\paragraph*{\textbf{\textit{Contributions}}}
Our approach is independent of isolation levels.
Nonetheless, it is particularly effective for strong isolation levels,
including \sser, \ser, and \si,
which impose higher execution overhead on databases
and involve greater verification complexity.
Our contributions are three-fold:

\begin{itemize}
	\item At the conceptual level,
	  we address inefficiencies in both history generation and verification stages
		during black-box checking of database isolation levels
		by proposing a unified approach centered on MTs.
	\item At the technical level, we propose the design of MTs,
		which are compact yet semantically rich,
		capable of capturing all 14 well-documented isolation anomalies
		specified in contemporary specification frameworks~\cite{Adya:PhDThesis1999,
			Framework:CONCUR2015, AnalysingSI:JACM2018, Complexity:OOPSLA2019, plume}.
		Furthermore, we develop a suite of highly efficient verification algorithms
		for strong isolation levels, leveraging the RMW pattern in MTs and unique values in MT histories.
		We also establish that verifying strong isolation levels
		for MT histories \emph{without} unique values is \nphard.
	\item At the practical level,
		we implement our verification algorithms,
		along with an MT workload generator, in a tool called \ourtool.
    Experimental results demonstrate that
		MTC outperforms state-of-the-art isolation checkers
		in the end-to-end checking process
		including both history generation and verification.
		\reviewright{R1}{O1}{purple}{Furthermore,
		we show that MTC can detect bugs
		across various isolation levels in production databases
		while maintaining the effectiveness of randomized testing
		with general workloads.}
\end{itemize}
\section{Background} \label{section:bg}

\subsection{Database Isolation Levels and Black-box Checking}
\label{ss:blackbox-checking}

Databases are essential to modern cloud-based systems and web applications,
serving as the backbone for geo-replicated data storage.
Applications coordinate their highly concurrent data accesses through transactions,
each composed of multiple read and write operations.
%
To balance data consistency with system performance,
databases provide various isolation levels.
We focus on strong isolation levels,
including \sser, \ser, and \si,
which are widely implemented in production databases~\cite{
	Fauna-ISO, MongoDB-ISO, PostgreSQL-ISO, MariaDB-ISO, YugabyteDB-ISO, TiDB-ISO}.
The gold standard, \ser, ensures that all transactions
appear to execute sequentially, one after another.
\si relaxes \ser but prevents undesirable anomalies
such as \fracturedread, \causalityviolation, and \lostupdate.
\sser strengthens \ser by enforcing the real-time order of transactions.

\begin{wrapfigure}{l}{3.8cm}
	\centering
	\includegraphics[width = 0.15\textwidth]{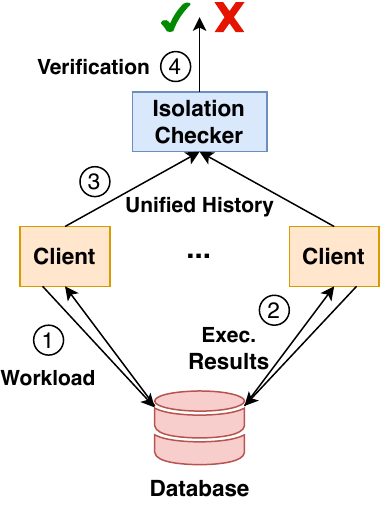}
	\caption{Workflow of black-box isolation checking.}
	\label{fig:black-box}
\end{wrapfigure}


Black-box database isolation checking involves two stages:
\emph{history generation} (Steps \circled{1}, \circled{2}, and \circled{3})
and \emph{history verification} (Step \circled{4});
see Figure~\ref{fig:black-box}.
First, clients send transactional requests to the database (\circled{1}),
treating it as a black-box system.
Each client records the requests sent to the database
along with the corresponding results received (\circled{2}).
For example, if a client performs a read operation,
it records both the requested object and the returned value.
The logs from all clients are combined into a single history,
which is provided to the isolation checker (\circled{3}).
Finally, the checker performs history verification,
deciding whether the history satisfies the specified isolation level.
If isolation violations are detected,
some checkers offer counterexamples (\circled{4}).

Following common practice in black-box isolation checking~\cite{Complexity:OOPSLA2019,PolySI:VLDB2023,Cobra:OSDI2020,plume,Viper:EuroSys2023},
we assume that every write operation
assigns a unique value to each object.
This guarantees that each read can be uniquely linked to
the transaction responsible for the corresponding write.
In practice, uniqueness can be easily enforced
by combining a client identifier with a local counter for each write.
\subsection{Objects and Operations} \label{ss:objects}

We consider a key-value database
that manages a set of objects $\Key = \set{x, y, \dots}$
associated with values from a set $\Val$.
Clients interact with the database
by issuing read and write operations on the objects,
grouped into transactions.
We denote by $\Op$ the possible operation invocations
in database executions:
$\Op = \set{\readevent(x, v), \writeevent(x, v) \mid x \in \Key, v \in \Val}$,
where $\readevent(x, v)$ represents a read operation
that reads value $v$ from object $x$, and
$\writeevent(x, v)$ represents a write operation
that writes value $v$ to object $x$.
\subsection{Relations and Orderings} \label{ss:relations}

A binary relation $R$ on a set $A$ is a subset of $A \times A$.
Let $I_{A} \triangleq \set{(a, a) \mid a \in A}$
be the identity relation on $A$.
For $a, b \in A$, we use $(a, b) \in R$ and $a \rel{R} b$ interchangeably.
The inverse $R^{-1}$ of a relation $R \subseteq A \times A$
is defined as $R^{-1} \triangleq \set{(b, a) \mid (a, b) \in R}$.
The reflexive closure $R?$ of a relation $R \subseteq A \times A$
is defined as $R? \triangleq R \cup I_{A}$.
A relation $R$ on a set $A$ is transitive
if $\forall a, b, c \in A.\; a \rel{R} b \land b \rel{R} c \implies a \rel{R} c$.
The transitive closure $R^{+}$ of a relation $R \subseteq A \times A$
is the smallest relation on $A$ that contains $R$ and is transitive.
A relation $R \subseteq A \times A$ is {\emph{acyclic}}
if $R^{+} \cap I_{A} = \emptyset$.
Given two binary relations $R, S \subseteq A \times A$,
we define their composition as
$R \comp S \triangleq \set{(a, c) \mid
  \exists b \in A.\; a \rel{R} b \land b \rel{S} c}$.
\reviewright{R1}{O6}{blue}{Note that
$R \comp S? = R \comp (S \cup I_{A}) = (R \comp S) \cup R$.}
A strict partial order is an irreflexive and transitive relation.
A strict total order is a relation that is a strict partial order and total.
We write $\_$ for a component that is irrelevant
and implicitly existentially quantified.
We use $\exists!$ to denote ``unique existence.''
\subsection{Histories and Dependency Graphs}
\label{ss:history-depgraph}

\begin{definition}[Transaction]
	\label{def:transaction}
	A {\it transaction} $T$ is a pair $(\opset, \po)$,
	where $\opset$ is a set of operations
	and $\po \subseteq \opset \times \opset$
	is a strict total order on $\opset$,
	called the {\it program order}.
\end{definition}

For a transaction $T$, we let $T \vdash \writeevent(x, v)$
if $T$ writes to $x$ and the last value written is $v$,
and $T \vdash \readevent(x, v)$
if $T$ reads from $x$ before writing to it
and $v$ is the value returned by the first such read.
We also use $\WriteTx_{x}$
to denote the set of transactions that write to $x$.

Transactions issued by clients are grouped into \emph{sessions},
where a session is a sequence of transactions.
We use a \emph{history} to record the client-visible results of such interactions.

\begin{definition}[History]
	\label{def:history}
	A history is a triple $\h = (\T, \SO, \RT)$,
	where $\T$ is a set of transactions,
	$\SO \subseteq \T \times \T$ is the {\it session order} on $\T$,
	and $\RT \subseteq \T \times \T$ is the {\it real-time order} on $\T$
	such that $\SO \subseteq \RT$
	and $T_{1} \rel{\RT} T_{2}$ if and only if
	$T_{1}$ finishes before $T_{2}$ starts in $\h$.
\end{definition}

The real-time order $\RT$ in the history is necessary for defining \sser,
which requires a transaction to observe the effects of all transactions
that finish before it starts~\cite{Spanner:TOCS2013}.
We only consider histories that satisfy
the \emph{internal consistency axiom}, denoted \intaxiom,
which ensures that, within a transaction,
a read from an object returns the same value
as the last write to or read from this object in the transaction.
Additionally, we assume that every history contains a special transaction $\bot_{T}$
which initializes all objects~\cite{AnalysingSI:JACM2018, Complexity:OOPSLA2019, MonkeyDB:OOPSLA2021},
unless explicitly specified otherwise.
This transaction precedes all the other transactions in the session order.

A \emph{dependency graph} extends a history with three relations,
i.e., $\WR$, $\WW$, and $\RW$,
representing three kinds of dependencies between transactions
in this history~\cite{Adya:PhDThesis1999, AnalysingSI:JACM2018}.
The $\WR$ relation associates a transaction
that reads some value with the one that writes this value.
The $\WW$ relation stipulates a strict total order
among the transactions on the same object.
The $\RW$ relation is derived from $\WR$ and $\WW$,
relating a transaction that reads some value to the one
that overwrites this value, in terms of the $\WW$ relation.

\begin{definition}[Dependency Graphs]
	\label{def:depgraph}
	A dependency graph is a tuple
	$\G = (\T, \SO, \RT, \WR, \WW, \WR)$
	where $(\T, \SO, \RT)$ is a history and
	\begin{itemize}
		\item $\WR: \Key \to 2^{\T \times \T}$ is such that
			\begin{itemize}
				\item $\forall x.\; \forall S \in \T.\;
					S \vdash \readevent(x, \_) \!\!\implies\!\! \exists!\; T \in \T.\; T \rel{\WR(x)} S$.
				\item $\forall x.\; \forall T, S \in \T.\;
					T \rel{\WR(x)} S \implies T \neq S \land
						\exists \valvar \in \Val.\; T \vdash \writeevent(x, \valvar) \land S \vdash \readevent(x, \valvar)$.
			\end{itemize}
		\item $\WW: \Key \to 2^{\T \times \T}$ is such that
			for every $x \in \Key$, $\WW(x)$ is a strict total order on the set $\WriteTx_{x}$;
		\item $\RW: \Key \to 2^{\T \times \T}$ is such that
			$\forall x.\; \forall T, S \in \T.\;
				T \rel{\RW(x)} S \iff T \neq S \land \exists T' \in \T.\; T' \rel{\WR(x)} T \land T' \rel{\WW(x)} S$.
  \end{itemize}
\end{definition}

We denote a component of $\G$, such as $\WW$, by $\WW_{\G}$.
\subsection{Characterizing Strong Isolation Levels}
\label{ss:strong-characterization-depgraph}

We review how the \sser, \ser, and \si{} isolation levels can be characterized
by dependency graphs that are acyclic or contain only specific cycles.
Specifically, a history $\H$ satisfies \sser{}, denoted $\H \models \sser$,
if and only if it satisfies \intaxiom{} and
one of its dependency graphs is acyclic.

\begin{definition}[Strict Serializability~\cite{Adya:PhDThesis1999, NCC:OSDI2023}]
	\label{def:sser}
	For a history $\H = (\T, \SO, \RT)$,
	\emph{\begin{align*}
    \H \models \sser & \iff \H \models \intaxiom \;\land \\
      &\exists \WR, \WW, \RW.\; \G = (\H, \WR, \WW, \RW) \;\land \\
      &\quad (\RT_{\G} \cup \WR_{\G} \cup \WW_{\G} \cup \RW_{G} \text{\it\; is acyclic}).
  \end{align*}}
\end{definition}

Being weaker than \sser{},
\ser{} does not require the preservation of the real-time order $\RT$
but only the session order $\SO$.
\begin{definition}[Serializability~\cite{AlgebraicLaw:CONCUR2017}]
	\label{def:ser}
	For a history $\H = (\T, \SO, \RT)$,
	\emph{\begin{align*}
    \H \models \ser & \iff \H \models \intaxiom \;\land \\
      &\exists \WR, \WW, \RW.\; \G = (\H, \WR, \WW, \RW) \;\land \\
      &\quad (\SO_{\G} \cup \WR_{\G} \cup \WW_{\G} \cup \RW_{G} \text{\it\; is acyclic}).
  \end{align*}}
\end{definition}

\si{} is even weaker than \ser:
a history satisifes \si{} if and only if
it satisfies \intaxiom{} and
one of its dependency graph (without the $\RT$ dependency edges) is acyclic
or contains only cycles with at least two adjacent $\RW$ edges. Formally,

\begin{definition}[Snapshot Isolation~\cite{AnalysingSI:JACM2018}]
	\label{def:si}
	For a history $\H = (\T, \SO, \RT)$,
    \emph{\begin{align*}
    \H \models \si & \iff \H \models \intaxiom \;\land \\
      &\exists \WR, \WW, \RW.\; \G = (\H, \WR, \WW, \RW) \;\land \\
      &\quad (((\SO_{\G} \cup \WR_{\G} \cup \WW_{\G}) \comp \RW_{\G}?) \text{\it\; is acyclic}).
  \end{align*}}
\end{definition}


\begin{figure}[t]
	\centering
	\includegraphics[width = 0.25\textwidth]{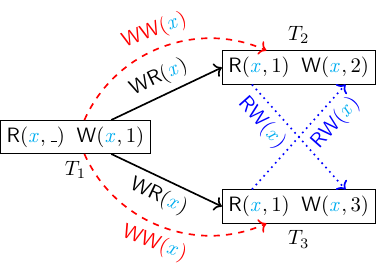}
	\caption{The \exception{} pattern:  $T_{2}$ and $T_{3}$ read the same value of $x$
		from $T_{1}$ and write different values.}
	\label{fig:pattern}
\end{figure}


\begin{example} \label{ex:si}
	Consider the history $\H$ of Figure~\ref{fig:pattern},
	where $T_{2}$ and $T_{3}$ read the same value of $x$
	from $T_{1}$ and then write different values to $x$.
	The $\WR$ and $\WW$ dependency edges from $T_{1}$ to $T_{2}$
	and $T_{3}$ respectively must be in any of the dependency graphs of $\H$.
	There are two cases regarding the $\WW$ dependency edge between $T_{2}$ and $T_{3}$.
	Suppose that $T_{2} \rel{\WW(x)} T_{3}$.
	Since $T_{1} \rel{\WW(x)} T_{2}$ and $T_{1} \rel{\WR(x)} T_{3}$,
	we have $T_{3} \rel{\RW(x)} T_{2}$.
	Therefore, the corresponding dependency graph contains a cycle
	$T_{2} \rel{\WW(x)} T_{3} \rel{\RW(x)} T_{2}$.
	By Definition~\ref{def:si},
	$\H$ does not satisfy \si.
	The case for $T_{3} \rel{\WW(x)} T_{2}$ is similar.
	Actually, the history $\H$ contains a \lostupdate anomaly.
\end{example}

\subsection{Linearizability} \label{ss:linearizability}

When each transaction comprises only one operation,
the \sser{} isolation level is the same as
the well-known \iso{Linearizability} (\lin) condition~\cite{Lin:TOPLAS1990}
for concurrent objects in shared-memory multiprocessors
and distributed systems~\cite{UnifiedTheory:JACM2004, HAT:VLDB2013}.
We use this connection when studying
the verification algorithm for lightweight-transaction histories
in Section~\ref{sss:alg-sser-cas}.

In the context of concurrent objects,
in addition to read and write operations,
we also consider the \emph{read\&write} operations~\cite{TSM:SIAM1997}.
A read\&write operation, denoted $\rwevent(x, v, v')$,
reads the value $v$ from object $x$ and then writes $v'$ to $x$.
From the perspective of executions,
a read\&write operation can be regarded as a transaction
that contains a read operation
followed by a write operation on the same object.
From the perspective of clients,
a read\&write operation invocation may correspond to
a lightweight transaction
(also known as Compare-And-Set (\cas) operation~\cite{TAOMP:Book2012})
that reads the current value of an object $x$ and,
if the value is equal to a given expected value $v$,
writes a new value $v'$ to the object.
If the value is not equal to $v$,
the lightweight transaction is equivalent to a simple read operation.
Another special case of lightweight transaction
is the \emph{insert-if-not-exists} operation~\cite{Cassandra-LWT},
which inserts an object with a value only if the object does not exist.
From the perspective of executions,
a successful {insert-if-not-exists} operation
is equivalent to a simple write operation.
Lightweight transactions have been widely adopted
in DBMSs like Apache Cassandra~\cite{Cassandra-LWT},
ScyllaDB~\cite{ScyllaDB-LWT},
PNUTS~\cite{PNUTS:VLDB2019},
Azure Cosmos DB~\cite{CosmosDB-LWT},
and etcd~\cite{etcd-LWT}.

When the start time $s$ and finish time $f$
of an operation is concerned,
we denote by $\readevent(s, f, x, v)$, $\writeevent(s, f, x, v)$,
and $\rwevent(s, f, x, v, v')$, correspondingly.
We use $o.s$ and $o.f$ to denote the start and finish time
of an operation $o$, respectively.
If an operation $o_{1}$ finishes before another operation $o_{2}$ starts,
then we say that $o_{1}$ precedes $o_{2}$ in the {real-time order}.
\lin{} requires that all operations
appear to be executed in some sequential order
that is consistent with the real-time order.

\begin{definition}[Linearizability~\cite{Lin:TOPLAS1990}]
	\label{def:linearizability}
	A history $\h$ is linearizable if and only if
	there exists a permutation $\Pi$ of all the operations in $\h$
	such that $\Pi$ preserves the real-time order of operations and
	follows the sequential semantics of each object.

\end{definition}

\begin{example} \label{ex:lin-cas}

\begin{figure}[t]
	\centering
	\begin{subfigure}[b]{0.42\columnwidth}
			\centering
			\includegraphics[width = \textwidth]{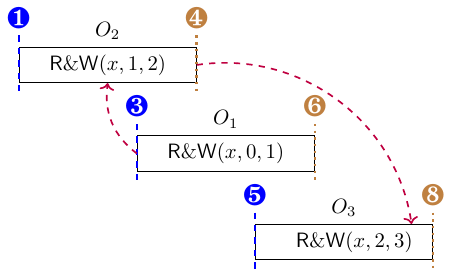}
			\caption{A linearizable history.}
			\label{fig:lin-rw}
	\end{subfigure}
	\begin{subfigure}[b]{0.47\columnwidth}
			\centering
			\includegraphics[width = \textwidth]{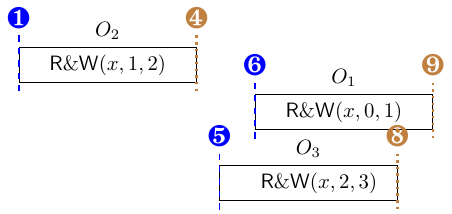}
			\caption{A non-linearizable history.}
			\label{fig:lin-rw-not}
	\end{subfigure}
	\caption{Illustration of linearizability on histories of \rwevent{} operations.
		The initial value of the object $x$ is $0$.}
	\label{fig:lin-cas}
\end{figure}

	Figure~\ref{fig:lin-rw} presents both a linearizable and
	a non-linearizable history of \rwevent{} operations.
	The history of Figure~\ref{fig:lin-rw} is linearizable
	as witnessed by the operation sequence consisting of $O_{1}$,
	$O_{2}$, and $O_{3}$ in this order.
	In contrast, the history of Figure~\ref{fig:lin-rw-not}
	is non-linearizable since $O_{1}: \rwevent(x, 0, 1)$ starts
	after $O_{2}: \rwevent(x, 1, 2)$ finishes.
\end{example}

\begin{figure*}[bp]
	\centering
	\begin{subfigure}[b]{0.18\textwidth}
		\centering
		\includegraphics[width = 1.00\textwidth]{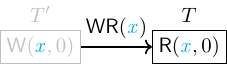}
		\caption{\thinairread.}
		\label{fig:thinairread}
	\end{subfigure}
	\hfill
	\begin{subfigure}[b]{0.25\textwidth}
		\centering
		\includegraphics[width = 1.00\textwidth]{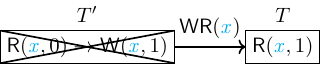}
		\caption{\abortedread.}
		\label{fig:abortedread}
	\end{subfigure}
	\hfill
	\begin{subfigure}[b]{0.18\textwidth}
		\centering
		\includegraphics[width = 0.80\textwidth]{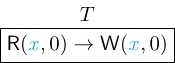}
		\caption{\futureread.}
		\label{fig:futureread}
	\end{subfigure}
	\hfill
	\begin{subfigure}[b]{0.30\textwidth}
		\centering
		\includegraphics[width = 1.00\textwidth]{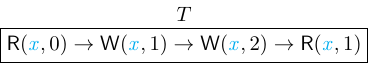}
		\caption{\notmylastwrite.}
		\label{fig:notmylastwrite}
	\end{subfigure}

	\vspace{5pt}
	\begin{subfigure}[b]{0.30\textwidth}
		\centering
		\includegraphics[width = 0.80\textwidth]{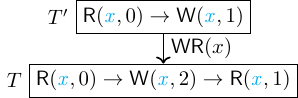}
		\caption{\notmyownwrite.}
		\label{fig:notmyownwrite}
	\end{subfigure}
	\hfill
	\begin{subfigure}[b]{0.32\textwidth}
		\centering
		\includegraphics[width = 0.80\textwidth]{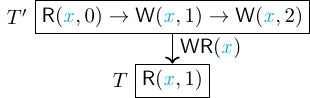}
		\caption{\intermediateread.}
		\label{fig:intermediateread}
	\end{subfigure}
	\hfill
	\begin{subfigure}[b]{0.35\textwidth}
		\centering
		\includegraphics[width = 0.90\textwidth]{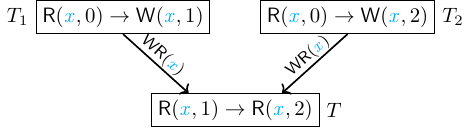}
		\caption{\nonrepeatableread.}
		\label{fig:nonrepeatableread}
	\end{subfigure}

	\vspace{5pt}
	\begin{subfigure}[b]{0.32\textwidth}
		\centering
		\includegraphics[width = \textwidth]{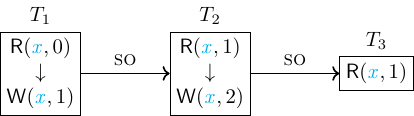}
		\caption{\sessionguaranteeviolation}
		\label{fig:session}
	\end{subfigure}
	\hfill
	\begin{subfigure}[b]{0.40\textwidth}
		\centering
		\includegraphics[width = \textwidth]{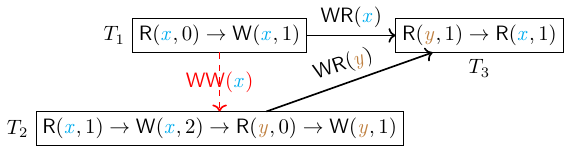}
		\caption{\nonmonoread}
		\label{fig:nonmonoread}
	\end{subfigure}
	\hfill
	\begin{subfigure}[b]{0.16\textwidth}
		\centering
		\includegraphics[width = \textwidth]{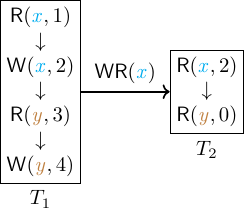}
		\caption{\fracturedread}
		\label{fig:fractured-reads}
	\end{subfigure}

	\vspace{5pt}
	\begin{subfigure}[b]{0.27\textwidth}
		\centering
		\includegraphics[width = \textwidth]{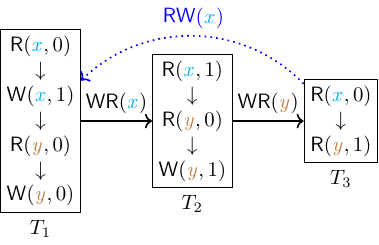}
		\caption{\causalityviolation}
		\label{fig:causality-violation}
	\end{subfigure}
  \hfill
	\begin{subfigure}[b]{0.18\textwidth}
		\centering
		\includegraphics[width = 0.80\textwidth]{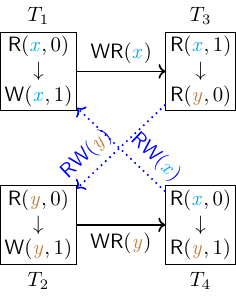}
		\caption{\longfork.}
		\label{fig:long-fork}
	\end{subfigure}
	\hfill
	\begin{subfigure}[b]{0.17\textwidth}
		\centering
		\includegraphics[width = \textwidth]{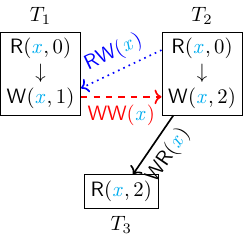}
		\caption{\lostupdate.}
		\label{fig:lost-update}
	\end{subfigure}
	\hfill
	\begin{subfigure}[b]{0.18\textwidth}
		\centering
		\includegraphics[width = \textwidth]{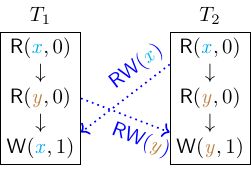}
		\caption{\writeskew.}
		\label{fig:write-skew}
	\end{subfigure}
	\caption{Capturing 14 isolation anomalies
	  in contemporary specification frameworks~\cite{Adya:PhDThesis1999,
		Framework:CONCUR2015, AnalysingSI:JACM2018, Complexity:OOPSLA2019, plume} by MTs.}
	\label{fig:anomalies}
\end{figure*}
\section{Mini-Transactions} \label{sec:mini}

In this section, we provide a precise definition of mini-transactions
and demonstrate that, despite their conceptual simplicity,
they can capture common data anomalies.

\begin{table*}[bp]
	\renewcommand{\arraystretch}{1.5}%
	\centering
	\caption{Common isolation anomalies captured by MTs
	  (corresponding to Figure~\ref{fig:anomalies}).}
	\label{table:anomalies}
	\resizebox{1.00\textwidth}{!}{%
	\begin{tabular}{|c|c|c|c|} \hline
		& {\bf Anomaly} & {\bf References} & {\bf Description}
	\\ \hline\hline
		(\subref{fig:thinairread}) & \thinairread & \cite{plume}
		& A transaction reads a value out of thin air.
	\\ \hline
		(\subref{fig:abortedread}) & \abortedread
		& \incell{c}{\cite[Phenomenon G1a]{Adya:PhDThesis1999} \\ \cite{plume}}
		& \incell{c}{A transaction reads a value from an aborted transaction.}
	\\ \hline
		(\subref{fig:futureread}) & \futureread
		& \cite{plume}
		& \incell{c}{A transaction reads from a write that occurs later in the same transaction.}
	\\ \hline
		(\subref{fig:notmylastwrite}) & \notmylastwrite
		& \cite{plume}
		& \incell{c}{A transaction reads from its own but not the last write on the same object.}
	\\ \hline
		(\subref{fig:notmyownwrite}) & \notmyownwrite
		& \cite{plume}
		& \incell{c}{A transaction does not read from its own write on the same object.}
	\\ \hline
		(\subref{fig:intermediateread}) & \intermediateread
		& \incell{c}{\cite[Phenomenon G1b]{Adya:PhDThesis1999} \\ \cite{plume}}
		& \incell{c}{A transaction reads a value that was later
			overwritten by the transaction that wrote it.}
	\\ \hline
		(\subref{fig:nonrepeatableread}) & \nonrepeatableread & \cite[Figure~3b]{Complexity:OOPSLA2019}
		& \incell{c}{A transaction reads multiple times
			from the same object but receives different values.}
	\\ \hline
		(\subref{fig:session}) & \sessionguaranteeviolation
		& \incell{c}{\cite[Figure~2a]{AnalysingSI:JACM2018} \\ {\cite[Figure~3c]{Complexity:OOPSLA2019}}}
		& \incell{c}{Transaction $T_{3}$ misses the effect of
		  the preceding transaction $T_{2}$ in the same session.}
		\\ \hline
		(\subref{fig:nonmonoread}) & \nonmonoread & \cite[Figure~3a]{Complexity:OOPSLA2019}
		& \incell{c}{Transaction $T_{3}$ reads $y$ from $T_{2}$
		  and {\it then} reads $x$ from $T_{1}$,
			but $T_{2}$ has overwritten $T_{1}$ on $x$.}
	\\ \hline
		(\subref{fig:fractured-reads}) & \fracturedread & \incell{c}{\cite[Figure~2c]{AnalysingSI:JACM2018} \\ {\cite[Figure~3d]{Complexity:OOPSLA2019}}}
		& \incell{c}{Transaction $T_{1}$ updates both $x$ and $y$,
		  but $T_{2}$ observes only the update to $x$.}
	\\ \hline
		(\subref{fig:causality-violation}) & \causalityviolation & \incell{c}{\cite[Figure~2d]{AnalysingSI:JACM2018} \\ {\cite[Figure~3e]{Complexity:OOPSLA2019}}}
		& \incell{c}{Transaction $T_{3}$ sees the effect of $T_{2}$ on $y$,
		  but misses the effect of $T_{1}$, which is seen by $T_{2}$, on $x$.}
	\\ \hline
		(\subref{fig:long-fork}) & \longfork & \incell{c}{\cite[Figure~2e]{AnalysingSI:JACM2018} \\ {\cite[Figure~3f]{Complexity:OOPSLA2019}}}
		& \incell{c}{Concurrent transactions $T_{1}$ and $T_{2}$
			write to $x$ and $y$, respectively. \\
			Transaction $T_{3}$ observes the write of $T_{1}$ to $x$
			but misses the write of $T_{2}$ to $y$, \\
			while $T_{4}$ observes the write of $T_{2}$ to $y$
			but misses the write of $T_{1}$ to $x$.}
	\\ \hline
		(\subref{fig:lost-update}) & \lostupdate & \incell{c}{\cite[Figure~2b]{AnalysingSI:JACM2018} \\ {\cite[Figure~3g]{Complexity:OOPSLA2019}}}
		& \incell{c}{Concurrent transactions $T_{1}$ and $T_{2}$
		  write to the same object,
			resulting in one of the writes getting lost.}
	\\ \hline
		(\subref{fig:write-skew}) & \writeskew & \incell{c}{\cite[Figure~2f]{AnalysingSI:JACM2018} \\ {\cite[Figure~3h]{Complexity:OOPSLA2019}}}
		& \incell{c}{Concurrent transactions $T_{1}$ and $T_{2}$
			read from both $x$ and $y$,
			and then write to $x$ and $y$, respectively.}
	\\ \hline
	\end{tabular}%
	}
\end{table*}

\subsection{Defining Mini-Transactions}

A mini-transaction is compact and adheres to the RMW pattern.

\begin{definition}[Mini-transaction; MT]
	A mini-transaction is a transaction
	meeting the following two criteria:
	\begin{itemize}
		\item It contains one or two read operations and at most two write operations.
		\item Each write operation, if \reviewleft{R1}{M1}{blue}{present},
		  is (not necessarily immediately)
			preceded by a read operation on the same object.
	\end{itemize}
\end{definition}

\begin{definition}[Mini-Transaction History]
	A mini-transaction history is a history
	comprising solely mini-transactions
	(except the initial transaction $\bot_{T}$)
	such that each write operation on the same object
	assigns a unique value.
\end{definition}

The first criterion of mini-transactions helps expedite database execution
during history generation,
while the second criterion, along with the unique values written,
guarantees the existence of efficient verification algorithms
for mini-transaction histories.
Note that according to this definition,
a read\&write operation can be regarded as a mini-transaction.
\subsection{Capturing Isolation Anomalies}
\label{ss:capturing-anomalies}

We demonstrate that mini-transactions are expressive enough
to characterize all common isolation anomalies.
As there is no standard for formally defining all kinds of isolation anomalies,
we refer to the state-of-the-art specification frameworks
for transactional isolation levels~\cite{Adya:PhDThesis1999,
  Framework:CONCUR2015, AnalysingSI:JACM2018, Complexity:OOPSLA2019, plume},
which, to some extent, have been shown to be equivalent.

Figure~\ref{fig:anomalies} illustrates 14 common isolation anomalies
specified in the frameworks above,
where each anomaly is represented by a mini-transaction history.
They cover both the intra-transactional anomalies
(i.e., Figure~\ref{fig:futureread}-\ref{fig:nonrepeatableread})
specified by \cite{Adya:PhDThesis1999, plume}
and the inter-transactional anomalies
(i.e., Figure~\ref{fig:session}-\ref{fig:write-skew})
emphasized by \cite{Framework:CONCUR2015, AnalysingSI:JACM2018, Complexity:OOPSLA2019}.
\reviewright{R1}{O4}{blue}{As demonstrated in Figure~\ref{fig:anomalies},
	a maximum of four operations per transaction
	is \emph{necessary and sufficient} for mini-transaction histories
	to capture these 14 isolation anomalies.
	Increasing the number of operations per transaction
	may incur a performance penalty during both history generation and verification.
}
Table~\ref{table:anomalies} lists these anomalies and their descriptions,
as well as the references to the frameworks that define them.
We believe this list is exhaustive with respect to
the aforementioned specification frameworks.

\section{Efficient History Verification Algorithms}
\label{section:algs}

Despite the \nphardness{} of verifying strong isolation levels
for general histories~\cite{SER:JACM1979, Complexity:OOPSLA2019},
we show that verifying whether an MT history of $n$ MTs
satisfies \sser, \ser, or \si{} can be remarkably efficient
in $O(n^2)$, $O(n)$, and $O(n)$ time, respectively,
thanks to the RMW patterns in MTs
and the unique value conditions in MT histories.
\reviewright{R3}{D4}{blue}{Our verification algorithms
are both sound and complete for a given MT history,
i.e., producing no false positives or negatives.}
Furthermore, in Appendix~\ref{section:appendix-complexity},
we establish that verifying strong isolation levels
for MT histories \emph{without} unique values is \nphard.

\subsection{Our Insight} \label{ss:idea}

The efficiency of our verification algorithms stems from a key insight:
\emph{the dependency graph of a mini-transaction history is {(nearly)} unique
and can be constructed efficiently}.
First, due to the unique values written,
the $\WR$ dependency in a mini-transaction history
is entirely determined.
Furthermore, since each write operation is preceded by a read operation
on the same object in mini-transactions,
the $\WW$ dependency is {\emph{almost}} determined.
\emph{Exceptions} only arise when multiple transactions
read the same value of an object and
subsequently write different values to it.
Finally, by definition, the $\RW$ dependency is entirely determined by
the $\WR$ and $\WW$ dependencies.
Hence, the challenge for the verification algorithms lies in
correctly handling the \emph{exceptions} in the $\WW$ dependency
for \sser, \ser, and \si, respectively.


We define the exceptions in the $\WW$ dependency
as the following \exception{} pattern,
which, as explained in Example~\ref{ex:si}, violates \si.

\begin{definition}[\exception{} Pattern]
	A history $\H$ contains an instance of the \exception{} pattern
	if there exist three transactions $T_{1}$, $T_{2}$, and $T_{3}$ in $\H$
	such that $T_{2}$ and $T_{3}$ read the same value of an object $x$ from $T_{1}$
	and then write different values to $x$.
\end{definition}

\begin{lemma} \label{lemma:fork-violates-si}
	If a history contains an instance of the \exception{} pattern,
	then it does not satisfy \si.
\end{lemma}



\subsection{Algorithms} \label{ss:algs}

The three verification algorithms for \sser, \ser, and \si{} respectively
follow the same structure, as outlined in Algorithm~\ref{alg:mini-checking}.\footnote{
  We assume that the input history
  $\H$ satisfies the $\intaxiom$ axiom (see Section~\ref{ss:history-depgraph}).
  In practice, we first check whether $\H$ contains
  $\thinairread$, $\abortedread$, or
  any intra-transactional anomalies (see Section~\ref{ss:capturing-anomalies}).}
They all construct the {(partial)} dependency graph $\G$
of the input history $\H$
based on the dependency relations $\RT$,
$\SO$, $\WR$, $\WW$, and $\RW$ by calling \consdep.
In \consdep, the $\RT$ dependency edges are added for \sser{} only
(line~\ref{line:rt-edges} of \consdep{}
and line~\ref{line:checksser-call-consdep} of \checksser).
Due to unique values written,
the $\WR$ dependency edges are entirely determined
(line~\ref{line:wr-edges} of \consdep).
Note that for mini-transaction histories,
the $\WW$ dependency edges are inferred from the $\WR$ dependency
(lines~\ref{line:ww-check}-\ref{line:ww-edges} of \consdep).
To make $\WW(x)$ transitive for each $x$,
\consdep{} then computes the transitive closure of $\WW(x)$ edges.
This step is convenient for the correctness proof
and can be optimized away (refer to Section~\ref{ss:opt}).
Finally, the $\RW$ dependency edges are added
based on the $\WR$ and $\WW$ dependencies
(lines~\ref{line:rw-dep}-\ref{line:rw-edges} of \consdep).


\begin{algorithm}[t]
  \small
  \caption{The algorithms for verifying \sser, \ser, and \si{} on mini-transaction histories.}
  \label{alg:mini-checking}
  \begin{algorithmic}[1]
    \Procedure{\consdep}{$\H, \rt$}
      \State $\G \gets (\T, \emptyset)$
        \Comment{initialize the dependency graph $\G$}
        \label{line:init-g}

      \hStatex
      \If{$\rt$}  \Comment{add $\RT$ edges for \checksser{} only}
        \ForAll{$T, S \in \T$ such that $T \rel{\RT} S$}
        \label{line:rt-dep}
        \State $\edges_{\G} \gets \edges_{\G} \cup \set{(T, S, \RT)}$
          \label{line:rt-edges}
        \EndFor
      \EndIf

      \ForAll{$T, S \in \T$ such that $T \rel{\SO} S$}
      \label{line:so-dep}
      \State $\edges_{\G} \gets \edges_{\G} \cup \set{(T, S, \SO)}$
        \label{line:so-edges}
      \EndFor

      \ForAll{$x \in \Key.\; T, S \in \T$ such that $T \rel{\WR(x)} S$}
        \label{line:wr-dep}
        \State $\edges_{\G} \gets \edges_{\G} \cup \set{(T, S, \WR(x))}$
          \label{line:wr-edges}
        \If{$\writeevent(x, \_) \in S$}
          \Comment{determine $\WW$ based on $\WR$}
          \label{line:ww-check}
          \State $\edges_{\G} \gets \edges_{\G} \cup \set{(T, S, \WW(x))}$
            \label{line:ww-edges}
        \EndIf
      \EndFor

      \hStatex
      \ForAll{$\optstyle{x \in \Key}$}
        \Comment{Convenient for the correctness proof
          and can be optimized away (see Section~\ref{ss:opt}).}
        \label{line:ww-opt-begin}
        \State $\optstyle{\text{compute the transitive closure of } \WW(x) \text{ edges in } \edges_{G}}$
        \label{line:ww-tc}
        \label{line:ww-opt-end}
      \EndFor
      \hStatex

      \ForAll{$x \in \Key.\; T', T, S \in \T$ such that $(T', T, \WR(x)) \in \edges_{G}
        \land (T', S, \WW(x)) \in \edges_{G}$}
          \label{line:rw-dep}
        \State $\edges_{\G} \gets \edges_{\G} \cup \set{(T, S, \RW(x))}$
          \label{line:rw-edges}
      \EndFor

      \State \Return $\G$
    \EndProcedure
  \end{algorithmic}

  \hStatex
  \begin{algorithmic}[1]
    \Procedure{\checksser}{$\H$}
      \label{line:proc-checksser}
      \State ${\G} \gets \Call{\consdep}{\H, \top}$
        \label{line:checksser-call-consdep}
      \State \Return $\Call{Acyclic}{\G}$
        \label{line:checksser-acyclic}
    \EndProcedure
  \end{algorithmic}

  \hStatex
  \begin{algorithmic}[1]
    \Procedure{\checkser}{$\H$}
      \label{proc:checkser}
      \State ${\G} \gets \Call{\consdep}{\H, \bot}$
        \label{line:checkser-call-consdep}
      \State \Return $\Call{Acyclic}{\G}$
        \label{line:checkser-acyclic}
    \EndProcedure
  \end{algorithmic}

  \hStatex
  \begin{algorithmic}[1]
    \Procedure{\checksi}{$\H$}
      \label{proc:checksi}
      \If{$\exists x \in \Key.\; \exists v, v' \neq v \in \Val.\;
        \exists T, S, S' \neq S \in \T.\;
        (T \rel{\WR(x)} S \land T \rel{\WR(x)} S') \land
        (S \vdash \writeevent(x, v) \land S' \vdash \writeevent(x, v'))$}
        \label{line:checksi-divergence}
        \State \Return \false
          \label{line:checksi-return-false}
          \Comment{{the \exception{} pattern (Figure~\ref{fig:pattern})}}
      \EndIf

      \State ${\G} \gets \Call{\consdep}{\H, \bot}$
        \label{line:checksi-call-consdep}

      \State $\G' \gets (V_{G}, (\SO_{\G} \cup \WR_{\G} \cup \WW_{\G}) \comp \RW_{\G}?)$
        \label{line:checksi-g-prime}
      \State \Return \Call{\acyclic}{$\G'$}
        \label{line:checksi-acyclic}
    \EndProcedure
  \end{algorithmic}
  \normalsize
\end{algorithm}

Subsequently, we check whether the partial dependency graph $\G$
is acyclic for \sser{} (line~\ref{line:checksser-acyclic} of \checksser)
and \ser{} (line~\ref{line:checkser-acyclic} of \checkser).
For the SI verification algorithm,
we check whether the induced graph
$\G' \gets (V_{\G}, (\SO_{\G} \cup \WR_{\G} \cup \WW_{G}) \comp \RW_{\G}?)$ of $\G$
is acyclic (line~\ref{line:checksi-acyclic} of \checksi).
Crucially, before checking the acyclicity of $\G'$,
we examine whether there exists a triple of transactions
$T$, $S$, and $S'$ such that $S$ and $S'$ read the same value
of some object from $T$ and
then write different values to this object
(line~\ref{line:checksi-divergence} of \checksi).
If this is the case, \checksi{} immediately return \false.

\begin{example}
  \label{ex:checkser-checksi}
  Consider the history $\H$ depicted in Figure~\ref{fig:pattern},
  which, according to Lemma~\ref{lemma:fork-violates-si},
  violates \si{} and \ser.
  Figure~\ref{fig:pattern} also illustrates the graph $\G$
  generated by \checkser{} and \checksi.
  Since $\G$ forms a cycle $T_{2} \rel{\RW(x)} T_{3} \rel{\RW(x)} T_{2}$,
  \checkser{} returns \false{} at line~\ref{line:checkser-acyclic},
  as expected.
  However, such a cycle is not forbidden by \si.
  Put differently, if we examine the induced graph $\G'$ of $\G$
  (line~\ref{line:checksi-g-prime} of \checksi),
  we would find it to be acyclic.
  \checksi{} effectively pinpoints the \si{} violation
  by detecting the \exception{} pattern
  early at line~\ref{line:checksi-return-false} in \checksi.
\end{example}

The correctness proofs of these three verification algorithms
are surprisingly intricate and are provided in
Appendix~\ref{section:appendix-proofs}.

\subsection{Optimizations} \label{ss:opt}


\begin{figure}[t]
	\centering
	\includegraphics[width = 0.30\textwidth]{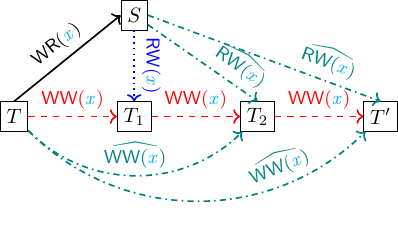}
	\caption{Illustrating the missing $\widehat{\WW}$
		and $\widehat{\RW}$ edges in $\widehat{\G}$.}
	\label{fig:ww-tc}
\end{figure}

In this section, we show that the step of
computing the transitive closure of the $\WW$ edges
at lines~\ref{line:ww-opt-begin}--\ref{line:ww-opt-end}
of \consdep{} can be optimized away
(the proof is given in Appendix~\ref{section:appendix-opt-proof}).
Let $\widehat{\G}$ be the graph constructed by \consdep{}
\emph{without} computing the transitive closure of the $\WW$ edges.
As illustrated in Figure~\ref{fig:ww-tc},
$\widehat{\G}$ may lack some $\widehat{\WW}$ edges
that can be derived from a sequence of $\WW$ edges by transitivity
and $\widehat{\RW}$ edges that are derived from $\widehat{\WW}$
edges and $\WR$ edges.

\begin{lemma}
	\label{lemma:widehat-rw}
	For each $S \rel{\widehat{\RW}} T'$ edge in $\G$,
	there exist a set of transactions $T_{1}, T_{2}, \ldots, T_{n}$
	such that $S \rel{\RW} T_{1} \rel{\WW} T_{2} \rel{\WW} \cdots
	  \rel{\WW} T_{n} \rel{\WW} T'$ in $\G$.
\end{lemma}


The key observation is that
any cycle in $\G$ can be transformed into a cycle in $\widehat{\G}$
by replacing the $\widehat{\WW}$ and $\widehat{\RW}$ edges
in the cycle with $\WW$ and $\RW$ edges in $\widehat{\G}$, respectively.
Therefore, we have the following theorem,
from which the correctness of the optimized versions of \checksser{}
and \checkser{}  follows directly.

\begin{theorem}
	\label{thm:G-Gbar-acyclic}
	$\G$ is acyclic if and only if $\widehat{\G}$ is acyclic.
\end{theorem}

	
			

Let $\widehat{\G}'$ be the induced graph of $\widehat{\G}$,
i.e., $\widehat{\G}' \gets (V_{\widehat{\G}},
  (\SO_{\widehat{\G}} \cup \WR_{\widehat{\G}} \cup \WW_{\widehat{G}})
	  \comp \RW_{\widehat{\G}}?)$.
The correctness of the optimized version of \checksi{}
follows from the following theorem.

\begin{theorem}
	\label{thm:G'-Gbar'-acyclic}
	$\G'$ is acyclic if and only if $\widehat{\G}'$ is acyclic.
\end{theorem}

\subsection{Time Complexity}
\label{ss:time-complexity}

Suppose that the input history comprises $n$ transactions,
and the dependency graph $\G$
returned by the (optimized) \consdep{} procedure contains $m$ edges.
Given that a mini-transaction may have
at most $n$ incoming $\RT$ edges,
one incoming $\SO$ edges
(note that, similar to the optimization in Section~\ref{ss:opt},
	the $\SO$ edges derived by transitivity can be optimized away),
two incoming $\WR$ edges,
two incoming $\WW$ edges, and two incoming $\RW$ edges,
we can deduce that $m = O(n)$.
Therefore, the time complexity of both the (optimized) \checkser{}
and \checksi{} procedures is $\Theta(n)$.
Moreover, since establishing the $\RT$ edges requires $\Theta(n^{2})$ time,
the time complexity of \checksser{} is $\Theta(n^{2})$.


\subsection{On \sser{} and Lightweight-transaction Histories} \label{sss:alg-sser-cas}

In this section, we show that
for lightweight-transaction histories with unique values,
we can check \sser{} more efficiently in (expected) linear time.
In such histories, each transaction is either a \emph{read\&write} operation
or an \emph{insert-if-not-exist} operation.
In lightweight-transaction histories,
we do \emph{not} assume the existence of an initial transaction $\bot_{T}$.
Instead, {insert-if-not-exists} operations
are utilized to insert objects with initial values.
It is important to note that in this context,
\sser{} degenerates to linearizability.
\subsubsection{Algorithm} \label{sss:alg-cas}


\begin{algorithm}[t]
  \caption{The algorithm for verifying linearizability
    on lightweight-transaction (LWT) histories.}
  \label{alg:sser-cas}
  \begin{algorithmic}[1]
    \Procedure{\vlcas}{$\H_{x}$}
      \Comment{$\H_{x}$ is non-empty} 
      \If{$\left\lvert\WriteTx_{x}\right\rvert \neq 1$}
        \label{line:vlcas-single-writer}
        \State \Return \false
        \Comment{{exactly one \emph{insert-if-not-exists}}}
      \EndIf

      \Statex \quad\;{\it \ding{182} construct the transaction chain if possible}
      \State $\valvar \gets \text{the value of the only write operation $\writeevent(x, v)$}$
        \label{line:vlcas-chain-begin}
      \State $\chain \gets \emptyseq$
        \Comment{initialize the chain to be empty}
      \For{$\T \neq \emptyset$} \label{alg-vvalid-test:loop}
        \If{{\small $\exists v' \in \Val.\; \exists!\; \nextinchain \in \T.\; \nextinchain = \rwevent(\_, \_, x, v, v')$}}
          \label{line:vlcas-chain-next}
          \State $\T \gets \T \setminus \set{\nextinchain}$
          \State $\chain \gets chain \circ \nextinchain$
            \Comment{append $\nextinchain$ to the chain}
          \State $\valvar \gets v'$
          \label{line:vlcas-chain-end}
        \Else
          \State \Return \false
        \EndIf
      \EndFor

      \Statex {\quad\; \it \ding{183} check the real-time requirement in linear time}
      \State $\earlycommittime \gets \infty$
        \label{line:vlcas-rt-begin}
      \For{$T \in \chain$ in reverse order}\label{alg-tvalid-test:loop}
        \If{$T.\starttime > \earlycommittime$}
          \State \Return \false
        \EndIf
        \State $\earlycommittime \gets \min\set{\earlycommittime, T.\committime}$
      \EndFor
      \State \Return \true
        \label{line:vlcas-rt-end}
    \EndProcedure
  \end{algorithmic}
\end{algorithm}

Since linearizability is a local property~\cite{Lin:TOPLAS1990},
meaning that a history $\H$ is linearizable
if and only if every sub-history restricted to individual object
is linearizable~\cite{Lin:TOPLAS1990},
Algorithm~\ref{alg:sser-cas} takes as an input
a history $\H_{x}$ of lightweight transactions on a single object $x$.
The history $\H_{x}$ is valid only if
it contains exactly one {insert-if-not-exists} operation on $x$
(line~\ref{line:vlcas-single-writer})
and the transactions in it can be organized into a chain
in which each transaction (i.e., an $\rwevent$ operation)
reads the value written by the previous transaction in the chain.
In the first step (\ding{182}), Algorithm~\ref{alg:sser-cas}
constructs this chain if possible
(lines~\ref{line:vlcas-chain-begin}--\ref{line:vlcas-chain-end}).
Then, in the second step (\ding{183}),
Algorithm~\ref{alg:sser-cas} checks the real-time requirement:
for each transaction, it \emph{cannot} start after
any of the following transactions in the chain commits.
\subsubsection{Efficiency} \label{sss:efficiency}

Consider a history $\H_{x}$ of $n$ lightweight transactions.
{The next transaction $t$ in the chain
(line~\ref{line:vlcas-chain-next}) can be found
in (expected) $O(1)$ time by using a hash table.}
Thus, step \ding{182} takes $O(n)$ time.
It is easy to see that step \ding{183} takes $O(n)$ time.
Therefore, Algorithm~\ref{alg:sser-cas} takes $O(n)$ time in total.

\section{Experiments} \label{section:experiments}

We have implemented our verification algorithms for MT histories
in a tool called \ourtool~\cite{MTC-Tool},
which incorporates three verification components for SSER, SER, and SI,
as well as a MT workload generator.
The implementation consists of a total of 800 lines of code in Java
and 300 lines of code in Go.

This section presents a comprehensive evaluation of \ourtool
and state-of-the-art black-box isolation checkers
that test databases under highly concurrent GT workloads.
We aim to address the following questions:

\begin{enumerate}
	\item[Q1:]
		How efficient are \ourtool's verification components,
		i.e., \ssertool, \sertool, and \sitool,
		compared to other checkers across different concurrency levels?
	\item[Q2:]
		How does \ourtool perform
		\reviewright{R4}{O3}{brown}{in terms of time and memory}
		in end-to-end checking when history generation is included?
	\item[Q3:]
		Is \ourtool's MT workload generator more effective,
		resulting in a higher success rate for committing transactions?
	\item[Q4:]
	  Can \ourtool effectively detect isolation violations
		in production databases?
\end{enumerate}

\usetikzlibrary{patterns}
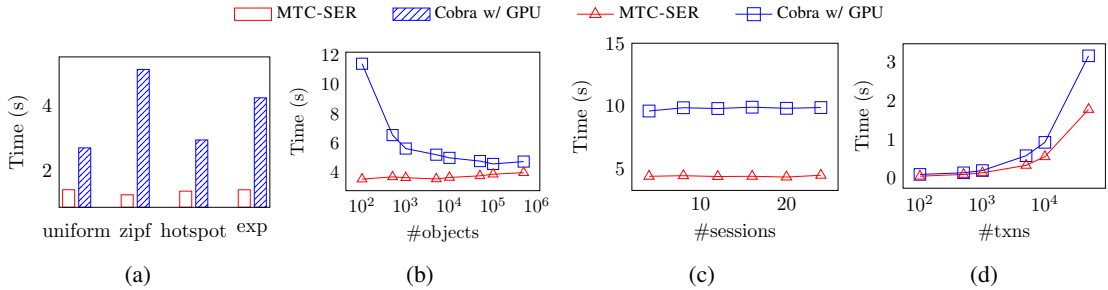
\begin{figure*}[htbp]
	\centering
	\resizebox{.50\textwidth}{!}{
		\begin{tikzpicture}
			\centering
			\begin{axis}[
				hide axis,
				width=2cm, 
				height=2cm, 
				legend style={at={(1,-0.1)},anchor=south, legend columns=4,draw=none},
				/tikz/every even column/.append style={column sep=0.5cm}
				]
				\addlegendimage{color=red, area legend}
				\addlegendentry{\footnotesize \sertool}
				\addlegendimage{color=blue,pattern=north east lines,pattern color=blue,area legend}
				\addlegendentry{\footnotesize Cobra w/ GPU}
				\addlegendimage{color=red,mark=triangle,mark size=3pt}
				\addlegendentry{\footnotesize \sertool}
				\addlegendimage{color=blue,mark=square,mark size=3pt}
				\addlegendentry{\footnotesize Cobra w/ GPU}
				\addplot[draw=none] coordinates {(0,0)}; 
			\end{axis}
		\end{tikzpicture}
	}

	\begin{subfigure}[b]{0.20\textwidth}
		\centering
		\includegraphics[width = \textwidth]{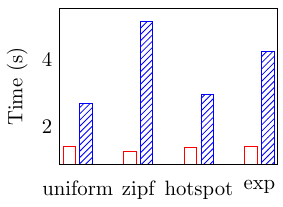}
		\caption{}
		\label{fig:ser-validation-dist}
	\end{subfigure}
	\begin{subfigure}[b]{0.20\textwidth}
		\centering
		\includegraphics[width = \textwidth]{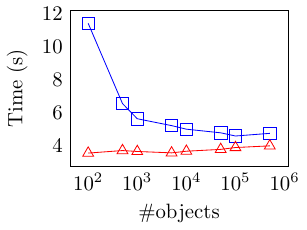}
		\caption{}
		\label{fig:ser-validation-keys}
	\end{subfigure}
	\begin{subfigure}[b]{0.20\textwidth}
		\centering
		\includegraphics[width = \textwidth]{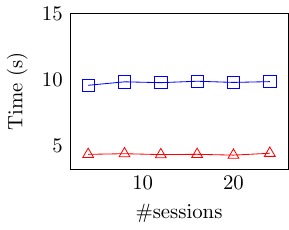}
		\caption{}
		\label{fig:ser-validation-sess}
	\end{subfigure}
	\begin{subfigure}[b]{0.20\textwidth}
		\centering
		\includegraphics[width = \textwidth]{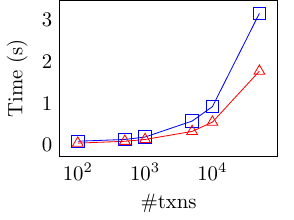}
		\caption{}
		\label{fig:ser-validation-txns}
	\end{subfigure}
	\caption{Performance comparison for verifying SER histories.}
	\label{fig:ser-validation}
\end{figure*}

\subsection{Workloads, Histories, and Setup}
\label{ss:setup}

\subsubsection{Workloads} \label{sss:workloads}

We generate two types of workloads: MT workloads and GT workloads.
MT workloads are used to compare \ourtool with existing checkers,
focusing on the efficiency of their verification components (Q1).
GT workloads, commonly employed by state-of-the-art checkers,
are used to evaluate end-to-end checking overhead,
highlighting the impact of MTs on the overall checking process (Q2).
Moreover, we assess the effectiveness of workload generators
by comparing their success rates in achieving committed transactions (Q3).
Finally, we evaluate the effectiveness of \ourtool
in detecting isolation violations in PostgreSQL and MongoDB (Q4).

Both workload generators are parametric
and designed to assess checker performance at various concurrency levels.
The MT workload generator parameters
include the number of sessions (\#sessions),
transactions (\#txns), objects (\#objects),
and the object-access distribution
(i.e., uniform, zipfian, hotspot, and exponential),
which defines workload skewness.
For the GT workload generator,
we use Cobra's implementation~\cite{Cobra:OSDI2020},
which allows configuration of \#objects, \#txns,
and \#ops/txn (operations per transaction).
Each {GT workload} consists of 20\% read-only transactions,
40\% write-only transactions, and 40\% RMW transactions.
During workload generation,
transactions are uniformly distributed across sessions,
and unique values are assigned to each object using counters.



\begin{figure*}[htbp]
	\centering
	\resizebox{0.40\textwidth}{!}{
		\begin{tikzpicture}
			\centering
			\begin{axis}[
				hide axis,
				width=2cm, 
				height=2cm, 
				legend style={at={(1,-0.1)},anchor=south, legend columns=4,draw=none},
				/tikz/every even column/.append style={column sep=0.5cm}
				]
				\addlegendimage{color=red,area legend}
				\addlegendentry{\footnotesize \sitool}
				\addlegendimage{color=blue,pattern=north east lines,pattern color=blue,area legend}
				\addlegendentry{\footnotesize PolySI}
				\addlegendimage{color=red,mark=triangle,mark size=3pt}
				\addlegendentry{\footnotesize \sitool}
				\addlegendimage{color=blue,mark=square,mark size=3pt}
				\addlegendentry{\footnotesize PolySI}
				\addplot[draw=none] coordinates {(0,0)}; 
			\end{axis}
		\end{tikzpicture}
	}

	\begin{subfigure}[b]{0.20\textwidth}
		\centering
		\includegraphics[width = \textwidth]{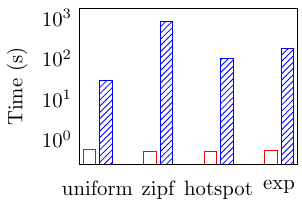}
		\caption{}
		\label{fig:si-validation-dist}
	\end{subfigure}
	\begin{subfigure}[b]{0.20\textwidth}
	\centering
	\includegraphics[width = \textwidth]{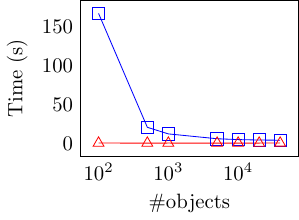}
	\caption{}
	\label{fig:si-validation-keys}
\end{subfigure}
	\begin{subfigure}[b]{0.20\textwidth}
		\centering
		\includegraphics[width = \textwidth]{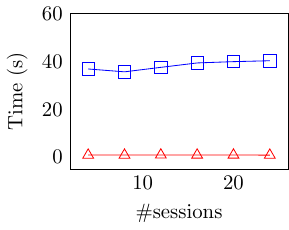}
		\caption{}
		\label{fig:si-validation-sess}
	\end{subfigure}
	\begin{subfigure}[b]{0.20\textwidth}
		\centering
		\includegraphics[width = \textwidth]{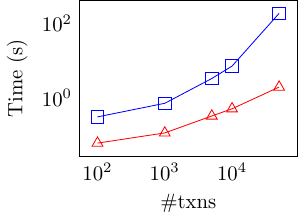}
		\caption{}
		\label{fig:si-validation-txns}
	\end{subfigure}
	\caption{Performance comparison for verifying SI histories.}
	\label{fig:si-validation}
\end{figure*}
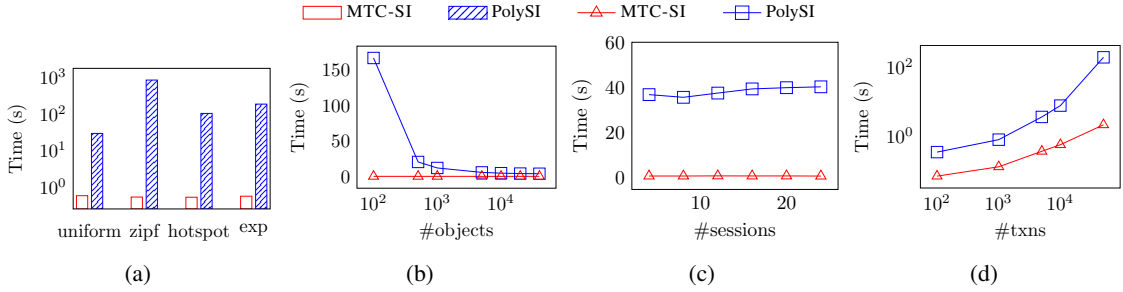
\subsubsection{Histories} \label{subsubsec:histories}

We use a PostgreSQL (v14.7) instance to
generate \textit{valid} histories without isolation violations.
Specifically, for \si, we set the isolation level to \iso{repeatable read}
(implemented as \si in PostgreSQL);
for \ser, we set the level to \iso{serializability}
to generate serializable histories.
To facilitate the comparison of SSER checkers
at varying concurrency levels,
we implement a parametric, synthetic history generator
tailored for lightweight transactions.\footnote{
	We observe that, for databases supporting lightweight transactions,
	e.g., Apache Cassandra,
	adjusting the workload parameters, as in our black-box setting,
	cannot predictably control the concurrency level in generated histories.}
{The generator's parameters include \#sessions,
\#txns/session (number of transactions per session),
and concurrent sessions
(the percentage of sessions issuing transactions concurrently).}

\subsubsection{Experimental Setup} \label{sss:setup}

We co-locate client/sessions and PostgreSQL on a local machine.
Each session issues a stream of transactions
generated by the workload generator to the database
and logs its execution history.
The histories from all sessions are then combined
and saved to a file,
which is used to benchmark the performance of the checkers.
All experiments are conducted on a system
with a 12-core CPU, 64GB of memory, and an NVIDIA P4 GPU.

\subsection{State-of-the-Art Isolation Checkers}

We consider the following checkers for comparison.
\begin{itemize}
	\item \textbf{SER}:
	  Cobra~\cite{Cobra:OSDI2020} is an efficient \ser checker
		that utilizes the advanced MonoSAT solver~\cite{MonoSAT:AAAI2015}.
		We also evaluate its GPU-accelerated version,
		serving as a strong baseline.
	\item \textbf{SI}:
	  PolySI~\cite{PolySI:VLDB2023} is an efficient SI checker
		that also relies on the MonoSAT solver.
 	\item \textbf{SSER}:
 	  Porcupine~\cite{porcupine} is a state-of-the-art LIN checker
		that does not rely on off-the-shelf solvers.
		It utilizes P-compositionality~\cite{horn2015faster}
		to further improve performance.
\end{itemize}

\reviewright{R3}{O1}{teal}{Both Cobra and PolySI transform the verification problem
	into a constraint-solving task,
	by constructing a polygraph from the history,
	where uncertainties in transaction execution order
	are represented as constraints.
	They prune constraints through domain-specific optimizations,
	encode the polygraph into SAT formulas,
	and use the MonoSAT solver to detect SER/SI-specific cycles.
	While effective for general workloads,
	we will shortly show that this approach is less efficient than
	our tailored verification algorithms on MT histories.
}


\begin{figure}
	\centering
	\resizebox{0.25\textwidth}{!}{
		\begin{tikzpicture}
			\centering
			\begin{axis}[
				hide axis,
				width=2cm, 
				height=2cm, 
				legend style={at={(1,-0.1)},anchor=south, legend columns=4,draw=none},
				/tikz/every even column/.append style={column sep=0.5cm}
				]
				\addlegendimage{color=blue,mark=square,mark size=2pt}
				\addlegendentry{\footnotesize Porcupine}
				\addlegendimage{color=red,mark=triangle,mark size=2pt}
				\addlegendentry{\footnotesize \ssertool}
				\addplot[draw=none] coordinates {(0,0)}; 
			\end{axis}
		\end{tikzpicture}
	}

	\begin{subfigure}{0.42\columnwidth}
		\includegraphics[width=\textwidth]{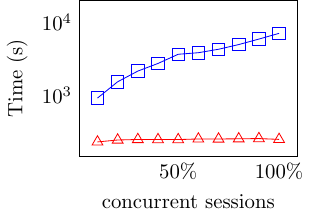}
		\caption{}
		\label{fig:lin-check-per-con}
	\end{subfigure}
	\begin{subfigure}{0.42\columnwidth}
		\includegraphics[width=\textwidth]{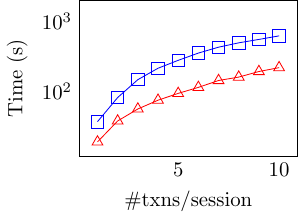}
		\caption{}
		\label{fig:lin-check-txn-per-clt}
	\end{subfigure}
	\caption{Performance comparison for verifying SSER histories. }
	\label{fig:sser-validation}
\end{figure}
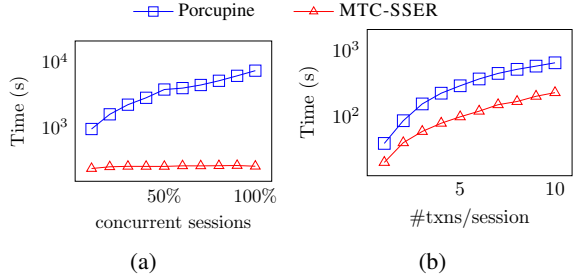

\subsection{Q1: History Verification Performance}
\label{ss:verification-performance}

\subsubsection{Serializability and Snapshot Isolation}
\label{sss:ser-si}

The first two sets of experiments compare \ourtool
with state-of-the-art \ser and \si checkers on MT histories.
The experimental results are presented
in Figures~\ref{fig:ser-validation} and~\ref{fig:si-validation}.

\sertool consistently outperforms Cobra's GPU-accelerated version
across various concurrency levels.
For MT histories, Cobra exhibits similar overhead
without GPU acceleration.
Under highly skewed object access patterns (implying high concurrency),
such as the zipfian distribution,
\sertool achieves approximately 5x better performance than Cobra
(Figure~\ref{fig:ser-validation-dist}).
Additionally, \sertool maintains stable performance
with respect to skewness,
while Cobra's verification time increases exponentially
with fewer objects (implying more skewed access patterns);
see Figure~\ref{fig:ser-validation-keys}.
Both checkers maintain stable performance with respect to the number of sessions,
yet \sertool consistently achieves around 2x better performance than Cobra
(Figure~\ref{fig:ser-validation-sess}).
Furthermore, {as the number of transactions increases},
Cobra's verification time escalates significantly faster
(Figure~\ref{fig:ser-validation-txns}).

Compared to PolySI in verifying SI,
\sitool demonstrates significantly greater performance gains
across various workloads.
For example, under the zipfian object-access distribution,
\sitool achieves approximately 1600x faster verification
(Figure~\ref{fig:si-validation-dist}).
With an increasing number of transactions,
\sitool reduces verification time by up to 93x
(Figure~\ref{fig:si-validation-txns}).



\subsubsection{Strict Serializability} \label{sss:sser-exp}

To benchmark the efficiency of SSER checkers,
we generate valid, synthetic SSER histories.
As shown in Figure~\ref{fig:sser-validation},
\ssertool outperforms Porcupine across various concurrency levels.
Under extreme concurrency
where all clients execute transactions simultaneously,
\ssertool demonstrates a substantial 28x improvement in verification.
Furthermore, \ssertool maintains stable performance
as the number of concurrent sessions increases
(Figure~\ref{fig:lin-check-per-con}).





\begin{figure*}[t]
	\centering
	\includegraphics[width = 0.75\textwidth]{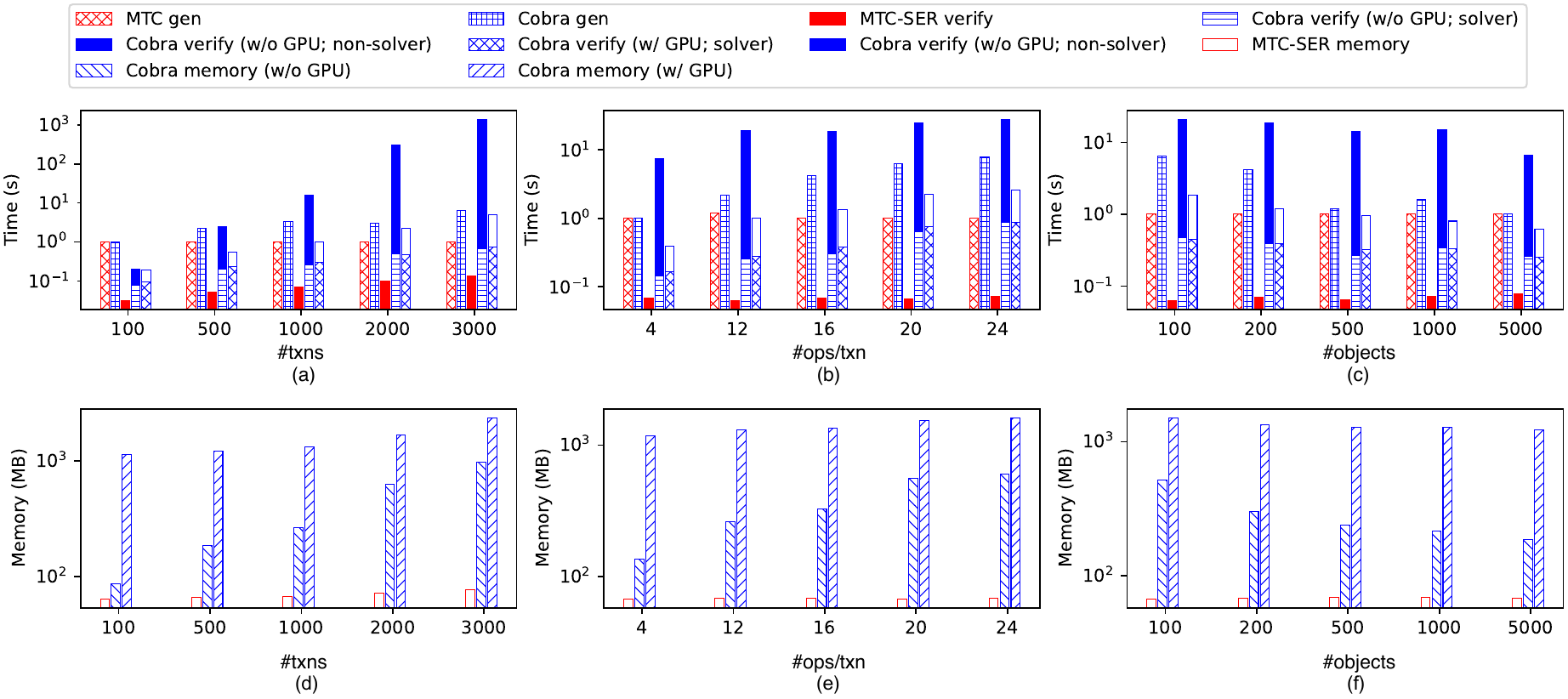}
	\caption{End-to-end checking performance,
	  with time decomposed into history generation and verification.}
	\label{fig:ser-e2e-time-memory-6subfigs}
\end{figure*}
\subsection{Q2: End-to-End Checking Performance}

To assess the performance improvement
that MTs bring to the end-to-end isolation checking process,
we compare baseline checkers using their standard GT workload generators.

As shown in Figures~\ref{fig:ser-e2e-time-memory-6subfigs}a-c,
\sertool with MT workloads substantially and consistently outperforms
Cobra with GT workloads in both history generation and verification.
achieving up to two and three orders of magnitude improvement
in history verification compared to Cobra with and without GPU, respectively.
\reviewleft{R1}{O3}{blue}{Figures~\ref{fig:ser-e2e-time-memory-6subfigs}a-c
also show that Cobra's non-solver components,
including polygraph construction, constraint pruning, and encoding,
are significantly more time-consuming than the solver itself.
This observation aligns with the findings
in the original works on Cobra~\cite{Cobra:OSDI2020}
and PolySI~\cite{PolySI:VLDB2023}.}
Additionally, as concurrency levels increase,
characterized by more transactions, more operations per transaction,
and less number of objects,
MT workload generation becomes increasingly more efficient compared to Cobra.
\reviewleft{R4}{O3}{brown}{Figures~\ref{fig:ser-e2e-time-memory-6subfigs}d-f show that
MTC-SER consistently consumes significantly less memory than Cobra,
achieving up to 30x and 14x improvement
compared to Cobra with and without GPU, respectively.}
We observe similar trends
\reviewleft{R4}{O3}{brown}{in both time and memory}
in the end-to-end isolation checking comparison
between \sitool and PolySI; see Appendix~\ref{section:appendix-si-e2e}.



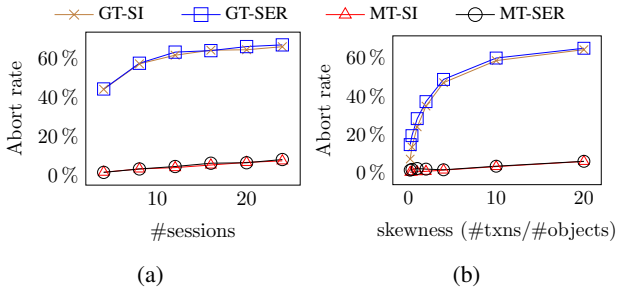
\begin{figure}[t]
	\centering
	\resizebox{.40\textwidth}{!}{
		\begin{tikzpicture}
			\centering
			\begin{axis}[
				hide axis,
				width=2cm, 
				height=2cm, 
				legend style={at={(1,-0.1)},anchor=south, legend columns=4,draw=none},
				/tikz/every even column/.append style={column sep=0.5cm}
				]
				\addlegendimage{color=brown,mark=x,mark size=3pt}
				\addlegendentry{\footnotesize GT-SI}
				\addlegendimage{color=blue,mark=square,mark size=3pt}
				\addlegendentry{\footnotesize GT-SER}
				\addlegendimage{color=red,mark=triangle,mark size=3pt}
				\addlegendentry{\footnotesize MT-SI}
				\addlegendimage{color=black,mark=o,mark size=3pt}
				\addlegendentry{\footnotesize MT-SER}
				\addplot[draw=none] coordinates {(0,0)}; 
			\end{axis}
		\end{tikzpicture}
	}
		\begin{subfigure}[b]{0.22\textwidth}
			\centering
			\includegraphics[width = \textwidth]{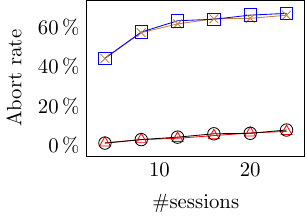}
			\caption{}
			\label{fig:abortion-sessions}
		\end{subfigure}
		\begin{subfigure}[b]{0.23\textwidth}
			\centering
			\includegraphics[width = \textwidth]{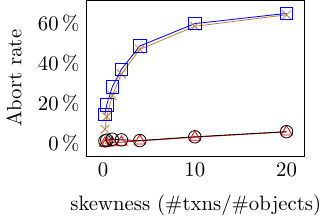}
			\caption{}
			\label{fig:abortion-skewness}
		\end{subfigure}
	\caption{Abort rates for GT and MT workloads. }
	\label{fig:abortion}
\end{figure}

\subsection{Q3: Effectiveness of Workload Generation} \label{subsec:abortion}

The effectiveness of stress-testing DBMSs
relies significantly on histories
containing numerous committed transactions.
Figure~\ref{fig:abortion} presents the abort rates
for executing MT and GT workloads in PostgreSQL under SER and SI.
For GT workloads, we use a moderate transaction size of 20 operations
(as we observed, larger transaction sizes lead to higher abort rates).

GT workloads result in substantially more aborted transactions,
leading to less effective histories.
As shown in Figure~\ref{fig:abortion-sessions},
even with only a few client sessions,
nearly half of the transactions are aborted.
The abort rate increases as more sessions are involved,
indicating greater stress on PostgreSQL.
Moreover, the GT workload generator is sensitive to access skewness.
In particular, when roughly 20 transactions access the same object,
over 60\% of transactions are aborted,
as shown in Figure~\ref{fig:abortion-skewness}.
In contrast, our MT workload generator
demonstrates robustness against variations in both cases.

\subsection{Q4: Detecting Isolation Bugs}
\label{subsec:detection}


\begin{figure}[t]
	\centering
	\begin{subfigure}[b]{0.24\textwidth}
		\centering
		\includegraphics[width=0.80\textwidth]{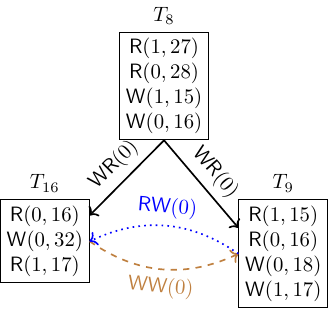}
		\caption{MariaDB Galera: \lostupdate that violates SI}
		\label{fig:bug-mariadb-lostupdate}
	\end{subfigure}
	\hfill
	\begin{subfigure}[b]{0.24\textwidth}
		\centering
		\includegraphics[width=0.80\textwidth]{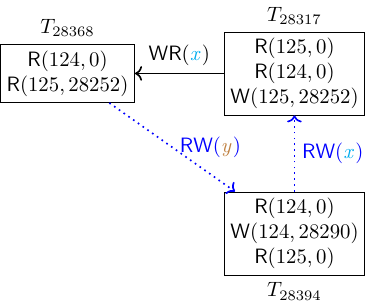}
		\caption{PostgreSQL: \writeskew that violates SER}
		\label{fig:bug-pg-writeskew}
	\end{subfigure}
	\caption{Rediscovered isolation bugs by \ourtool.}
	\label{fig:bugs}
\end{figure}

\subsubsection{Rediscovering Bugs} \label{sss:rediscovering-bugs}

\ourtool can successfully (re)discover real-world bugs
across different isolation levels with MTs,
as summarized in Table~\ref{table:bugs}
and illustrated in Figure~\ref{fig:bugs}.
These bugs appear in six releases of five databases
and represent various data anomalies,
e.g., the \lostupdate anomaly found in MariaDB Galera (Figure~\ref{fig:bug-mariadb-lostupdate})
which violates the claimed SI.
\ourtool demonstrates high efficiency in detecting these bugs,
particularly with instant history verification.
Figure~\ref{fig:bug-pg-writeskew} depicts
the \writeskew anomaly found in PostgreSQL,
which violates the claimed SER.
\sertool reports this cycle containing two consecutive $\RW$ edges (see also Figure~\ref{fig:write-skew}),
where each involved MT has only two or three operations.
We defer other bugs found by \ourtool to Appendix~\ref{section:appendix-bugs}.
Notably, the counterexamples returned by \ourtool are relatively concise
and easy to interpret with MTs.


\renewcommand*{\arraystretch}{1.2}
\begin{table*}[t]
	\centering
	\caption{Summary of rediscovered isolation bugs.
	  Counterexample (CE) position refers to the position of the first MT
		of the counterexample in the history.
		Time includes both history generation (Hist. Gen.) and verification (Hist. Verify).}
	\label{table:bugs}
	\resizebox{0.85\textwidth}{!}{
		\begin{tabular}{cllccrrc}
			\hline
			\textbf{Isolation Level} & \textbf{Anomaly}  & \textbf{Database} &
			\textbf{{Bug Report}} & \textbf{{Status}} &
			\textbf{CE Position}  & \textbf{Hist. Gen.} &  \textbf{Hist. Verify}  \\
			\hline
			SI & \lostupdate & MariaDB Galera 10.7.3 & \cite{mariadb-issue} & Fixed & 20 & $<$ 1s  & $<$ 1s\\
			SI & \abortedread & MongoDB 4.2.6 & \cite{MongoDB-Jepsen} & Fixed & 0.3k  & 123s & $<$ 1s\\
			SI & \causalityviolation & Dgraph 1.1.1 & \cite{Dgraph-Jepsen} & Confirmed & 2.5k  & 615s  & $<$ 1s\\

			SER & \writeskew & PostgreSQL 12.3 & \cite{pg-issue} & Fixed & 3.5k  & 120s & $<$ 1s\\
			SER & \longfork & PostgreSQL 11.8 & \cite{jepsen-pg} & Fixed & 24k  & 240s & $<$ 2s\\

			SSER & \abortedread & Apache Cassandra 2.0.1 & \cite{jepsen-cassandra} & Fixed & 30 & 60s & $<$ 1s\\
			\hline
		\end{tabular}
	}
\end{table*}



\begin{figure}[t]
	\centering
	\begin{subfigure}[b]{0.24\textwidth}
		\centering
		\includegraphics[width = 1.00\textwidth]{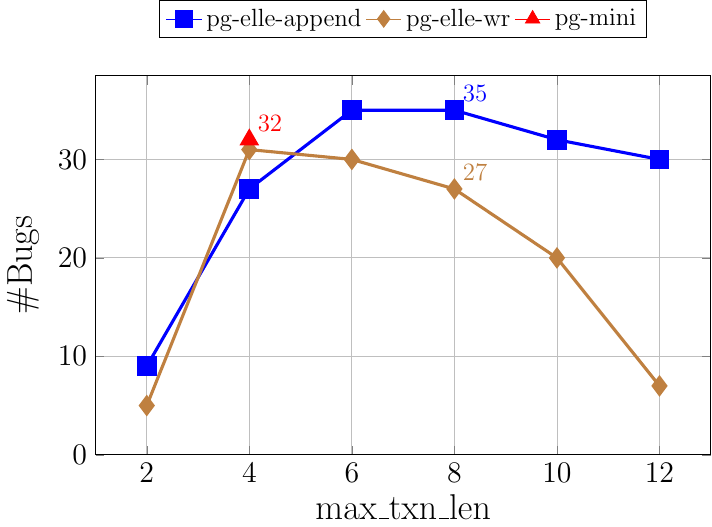}
		\caption{On PostgreSQL and SER.}
		\label{fig:effectiveness-pg}
	\end{subfigure}
	\hfill
	\begin{subfigure}[b]{0.24\textwidth}
		\centering
		\includegraphics[width = 1.00\textwidth]{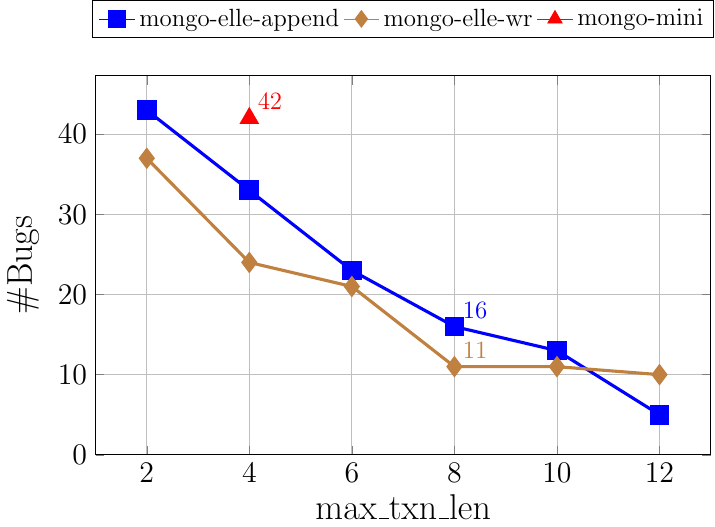}
		\caption{On MongoDB and SI}
		\label{fig:effectiveness-mongodb}
	\end{subfigure}
	\caption{Effectiveness of MTC in detecting isolation bugs.}
	\label{fig:effectiveness}
\end{figure}

\begin{figure}[t]
	\centering
	\begin{subfigure}[b]{0.24\textwidth}
		\centering
		\includegraphics[width = 1.00\textwidth]{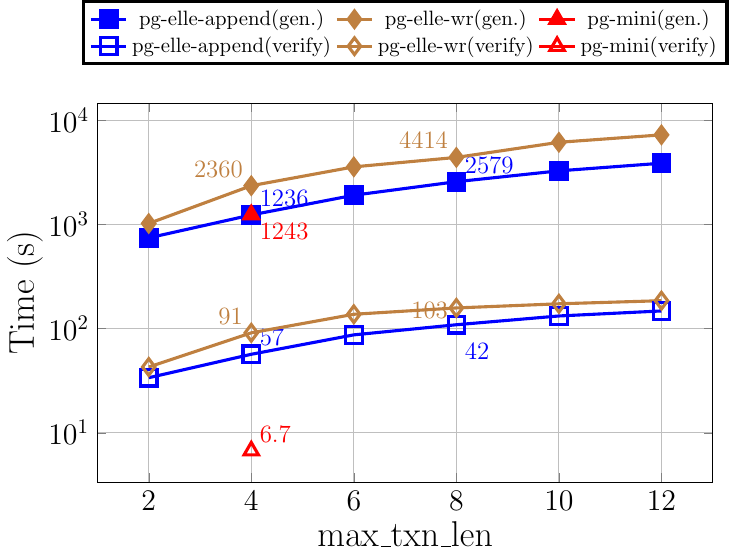}
		\caption{On PostgreSQL and SER.}
		\label{fig:mt-elle-pg}
	\end{subfigure}
	\hfill
	\begin{subfigure}[b]{0.24\textwidth}
		\centering
		\includegraphics[width = 1.00\textwidth]{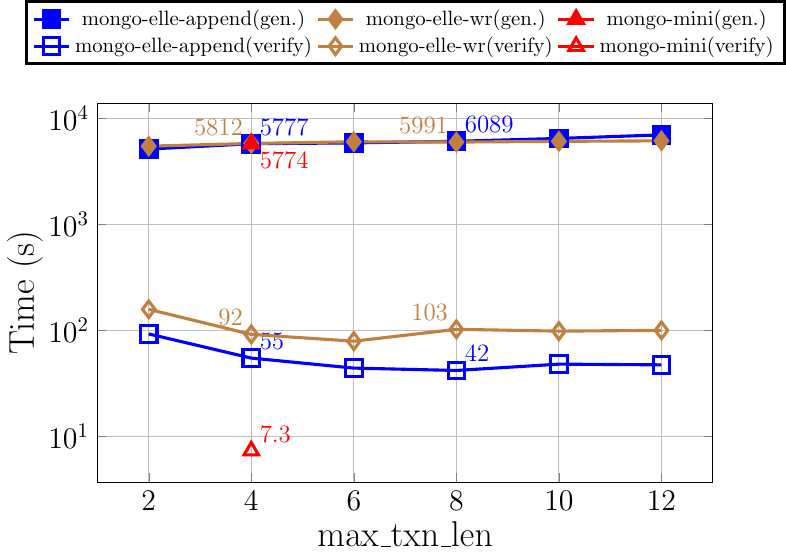}
		\caption{On MongoDB and SI}
		\label{fig:mt-elle-mongodb}
	\end{subfigure}
	\caption{End-to-end checking time for \ourtool and Elle.}
	\label{fig:mt-elle}
\end{figure}
\subsubsection{Effectiveness} \label{sss:effectiveness}

\reviewright{R1}{O1}{purple}{To evaluate the effectiveness of \ourtool
in detecting isolation bugs,
we compare it against randomized testing of PostgreSQL (v12.3)
and MongoDB (v4.2.6) using Elle~\cite{Elle:VLDB2020} under GT workloads,
including list append and read-write registers,
with varying transaction sizes.
Every database execution is required to successfully commit 3,000 transactions.
To increase concurrency, we set the number of objects to 10
and use an ``exponential'' object-access distribution.
For each configuration, the experiment lasts 30 minutes.}

\reviewleft{R1}{O5}{blue}{In Figure~\ref{fig:effectiveness},
the $x$-axis labeled ``max\_txn\_len''
represents the \emph{maximum} number of operations per transaction.
For MTC, it is 4.}
\reviewright{R4}{O2}{brown}{We use ``pg'' for PostgreSQL
and ``mongo'' for MongoDB,
and ``append'', ``wr'', and ``mini'' for the list append, read-write registers,
and MT workloads, respectively.}
\reviewright{R1}{O1}{purple}{
	As shown in Figure~\ref{fig:effectiveness-pg},
	within 30 minutes, MTC detects bugs in 32 successful trials on PostgreSQL,
	while Elle achieves the highest bug detection rate
	with 35 successful trials at a maximum transaction length of 8 operations
	under the list append workload.
	On the other hand, MTC consistently detects bugs
	in more successful trials than Elle under the read-write register workload.
	On MongoDB (Figure~\ref{fig:effectiveness-mongodb}),
	MTC is also highly competitive in bug detection compared to Elle.
}

Note that Elle's effectiveness in bug detection
is highly sensitive to transaction sizes across databases,
whereas a transaction size of 4 serves as a baseline.

\reviewleft{R3}{O2}{teal}{
	We also report the average time
	for history generation and verification for MTC and Elle.
	Figure~\ref{fig:mt-elle} shows that
	MTC consistently outperforms Elle in both tasks.
	For example, on PostgreSQL with ``max\_txn\_len = 8''
	under a read-write register workload,
	MTC achieves up to 3.5x and 15x speedups
	in history generation and verification, respectively.
}

\section{Related Work}
\label{sec:related}

We focus our comparison on the state-of-the-art black-box checkers for  isolation levels, 
excluding white-box approaches or those targeting database protocols~\cite{ConsMaude:TACAS2019,Emme:EuroSys2024,VerIso}.

A recent advance leverages SMT solvers
to check strong database isolation levels.
Representative checkers include
Viper~\cite{Viper:EuroSys2023}
and PolySI~\cite{PolySI:VLDB2023} for \si,
as well as Cobra~\cite{Cobra:OSDI2020} for \ser.
Each checker encodes a database execution history
as a \emph{polygraph}~\cite{SER:JACM1979} or its variant
and
prunes constraints through domain-specific optimizations.
They all invoke the off-the-shelf MonoSAT~\cite{MonoSAT:AAAI2015} solver
tailored for checking graph properties like acyclicity.
Our \ourtool   also tests DBMSs with randomly generated, highly concurrent workloads.
However, it substantially improves efficiency in both stages of history generation and verification via MTs.
Upon detecting a bug,
Cobra and Viper  return unsatisfied clauses as the counterexample.

Isolation checkers not based on SMT solving
employ graph traversal algorithms
to identify isolation bugs 
in a dependency graph.
 The dbcop tool~\cite{Complexity:OOPSLA2019}
builds on a polynomial-time algorithm for verifying \ser (with a fixed number of sessions)  and a polynomial-time algorithm for reducing verifying \si to verifying \ser.
However, it is less efficient than Cobra and PolySI~\cite{Cobra:OSDI2020,PolySI:VLDB2023}.
Moreover, when identifying a bug,
dbcop only returns ``false'', without providing any details.

Elle~\cite{Elle:VLDB2020} is an integrated isolation checker of
 the Jepsen
testing framework~\cite{jepsen}.
It leverages Jepsen's \emph{list-append} workloads
to efficiently infer write-write dependencies.
Specifically, reading a list of $n$ values infers
a potentially sequential version order
among the corresponding $n$ append operations.
Our utilization of the RMW pattern shares the similar spirit:
we can build a chain of MTs with consecutive writes on the same object. 
Elle also incorporates a linearizability checker called Knossos~\cite{Knossos},
which operates without using an off-the-shelf solver.
However, it is less efficient than Porcupine~\cite{porcupine}.

Porcupine is a fast LIN checker
for verifying histories on concurrent data types.
It implements \emph{P-compositionality}~\cite{horn2015faster},
a generalization of the locality principle~\cite{Lin:TOPLAS1990} for LIN.
Our \ssertool checker concentrates on histories of read\&write operations
on a single read-write register (owing to the locality principle).
By initially establishing a chain of these read\&write operations
(owing to the RMW pattern)
and then verifying the real-time requirement,
it achieves an 
$O(n)$ time complexity
for histories comprising $n$ operations.

Plume~\cite{plume}  is tailored for checking weak
isolation levels, including \iso{transactional causal consistency} and \iso{read committed}.
It builds on modular transactional anomalous patterns that faithfully characterize these isolation levels.
By utilizing vectors and tree clocks, Plume achieves superior performance in checking weak isolation levels.
This work complements Plume
by optimizing checking performance for stronger isolation levels.

TxCheck~\cite{txcheck} uses complex SQL queries
to detect logic bugs in database transactions,
which are not necessarily isolation bugs.
This approach enables the testing of advanced database features,
such as range queries and indexes,
complementing existing checkers that are primarily limited to simpler data models,
like read-write registers.
However, TxCheck relies on metamorphic testing~\cite{Metamorphic:arXiv2020}
to construct its test oracles,
which cannot establish the ground truth.
Consequently, the observed ``incorrect'' query results
may include false positives.
Furthermore, as TxCheck is not specifically designed
for discovering isolation bugs,
it may generate false positives
due to spurious dependencies created between SQL statements.

\section{Conclusion and Future Work}
\label{section:conclusion}

Utilizing MTs, we have tackled the inefficiency issues throughout the end-to-end black-box checking process for three strong  isolation levels in a unified manner.
Moreover, we have demonstrated the  effectiveness of our \ourtool tool
in detecting isolation bugs with MT workloads.

\reviewright{R1}{O1}{purple}{
	It remains to explore MTC's effectiveness with respect to
  workload parameters such as the
  object-access distribution and the number of objects.}
\reviewright{R1}{O2}{blue}{We also plan to
	evaluate MTC's performance in verifying SSER on arbitrary MT histories
	(besides LWT histories)
	by conducting experiments on FaunaDB
	and VoltDB~\cite{FaunaDB-Jepsen, VoltDB-Jepsen}.
	A key challenge is to accurately collect transaction
	start and finish wall-clock timestamps,
	while handling potential clock skew
	to minimize false positives and negatives.}
\reviewright{R3}{D3}{brown}{Furthermore,
  extending MTC for
	verifying weaker isolation levels
	would broaden its applicability.}

\reviewright{R3}{D2}{teal}{Our MTC tool includes a workload generator
that creates randomized test cases with MT histories.
We believe that guiding the generator to cover the isolation anomalies
depicted in Figure~\ref{fig:anomalies},
as shown in \cite{plume,DBStorm:ISSTA2024},
would further benefit developers.}
\reviewright{R3}{O3}{violet}{Moreover, an interesting avenue for future work
	is to explore how to execute MTs in a controlled manner,
	such as simulating stalls, altering state visibility, and injecting faults,
	to trigger specific anomalies.
	Finally, we plan to integrate the MTC tool into our IsoVista checking system~\cite{IsoVista},
	complementing its support for checking-as-a-service.}

\section*{Acknowledgments}
We thank the anonymous reviewers for their valuable feedback.
This work is supported by NSFC (62472214),
Natural Science Foundation of Jiangsu Province (BK20242014),
and ETH Zurich Career Seed Award.
Yuxing Chen is the corresponding author.

\newpage

\balance
\bibliographystyle{IEEEtran}
\bibliography{ref}

\begin{thebibliography}{10}
\providecommand{\url}[1]{#1}
\csname url@samestyle\endcsname
\providecommand{\newblock}{\relax}
\providecommand{\bibinfo}[2]{#2}
\providecommand{\BIBentrySTDinterwordspacing}{\spaceskip=0pt\relax}
\providecommand{\BIBentryALTinterwordstretchfactor}{4}
\providecommand{\BIBentryALTinterwordspacing}{\spaceskip=\fontdimen2\font plus
\BIBentryALTinterwordstretchfactor\fontdimen3\font minus \fontdimen4\font\relax}
\providecommand{\BIBforeignlanguage}[2]{{%
\expandafter\ifx\csname l@#1\endcsname\relax
\typeout{** WARNING: IEEEtran.bst: No hyphenation pattern has been}%
\typeout{** loaded for the language `#1'. Using the pattern for}%
\typeout{** the default language instead.}%
\else
\language=\csname l@#1\endcsname
\fi
#2}}
\providecommand{\BIBdecl}{\relax}
\BIBdecl

\bibitem{TIS:Book2001}
G.~Weikum and G.~Vossen, \emph{Transactional Information Systems: Theory, Algorithms, and the Practice of Concurrency Control and Recovery}.\hskip 1em plus 0.5em minus 0.4em\relax San Francisco, CA, USA: Morgan Kaufmann Publishers Inc., 2001.

\bibitem{CritiqueANSI:SIGMOD1995}
H.~Berenson, P.~Bernstein, J.~Gray, J.~Melton, E.~O'Neil, and P.~O'Neil, ``A critique of {ANSI SQL} isolation levels,'' in \emph{SIGMOD'95}.\hskip 1em plus 0.5em minus 0.4em\relax ACM, 1995, pp. 1--10.

\bibitem{Elle:VLDB2020}
K.~Kingsbury and P.~Alvaro, ``{Elle}: Inferring isolation anomalies from experimental observations,'' \emph{Proc. VLDB Endow.}, vol.~14, no.~3, pp. 268--280, Nov. 2020.

\bibitem{PolySI:VLDB2023}
K.~Huang, S.~Liu, Z.~Chen, H.~Wei, D.~A. Basin, H.~Li, and A.~Pan, ``Efficient black-box checking of snapshot isolation in databases,'' \emph{Proc. {VLDB} Endow.}, vol.~16, no.~6, pp. 1264--1276, 2023.

\bibitem{Complexity:OOPSLA2019}
R.~Biswas and C.~Enea, ``On the complexity of checking transactional consistency,'' \emph{Proc. ACM Program. Lang.}, vol.~3, no. OOPSLA, 2019.

\bibitem{plume}
S.~Liu, L.~Gu, H.~Wei, and D.~Basin, ``Plume: Efficient and complete black-box checking of weak isolation levels,'' \emph{Proc. {ACM} Program. Lang.}, vol.~8, no. {OOPSLA2}, 2024.

\bibitem{txcheck}
Z.-M. Jiang, S.~Liu, M.~Rigger, and Z.~Su, ``Detecting transactional bugs in database engines via graph-based oracle construction,'' in \emph{OSDI'23}.\hskip 1em plus 0.5em minus 0.4em\relax USENIX Association, 2023, pp. 397--417.

\bibitem{jepsen-pg}
{Kyle Kingsbury}, ``Jepsen testing of {PostgreSQL} 12.3,'' \url{https://jepsen.io/analyses/postgresql-12.3}, Accessed in February, 2024.

\bibitem{Cobra:OSDI2020}
C.~Tan, C.~Zhao, S.~Mu, and M.~Walfish, ``{COBRA}: Making transactional key-value stores verifiably serializable,'' in \emph{OSDI'20}, 2020.

\bibitem{Viper:EuroSys2023}
J.~Zhang, Y.~Ji, S.~Mu, and C.~Tan, ``Viper: A fast snapshot isolation checker,'' ser. EuroSys'23.\hskip 1em plus 0.5em minus 0.4em\relax ACM, 2023, p. 654–671.

\bibitem{porcupine}
A.~Athalye, ``Porcupine: A fast linearizability checker in {Go},'' \url{https://github.com/anishathalye/porcupine}, Accessed in December, 2023.

\bibitem{Adya:PhDThesis1999}
A.~Adya, ``Weak consistency: A generalized theory and optimistic implementations for distributed transactions,'' Ph.D. dissertation, Massachusetts Institute of Technology, USA, 1999.

\bibitem{AnalysingSI:JACM2018}
A.~Cerone and A.~Gotsman, ``Analysing snapshot isolation,'' \emph{J. ACM}, vol.~65, no.~2, 2018.

\bibitem{HAT:VLDB2013}
P.~Bailis, A.~Davidson, A.~Fekete, A.~Ghodsi, J.~M. Hellerstein, and I.~Stoica, ``Highly available transactions: Virtues and limitations,'' \emph{Proc. VLDB Endow.}, vol.~7, no.~3, pp. 181--192, nov 2013.

\bibitem{noc-noc}
S.~Liu, L.~Multazzu, H.~Wei, and D.~A. Basin, ``{NOC-NOC}: Towards performance-optimal distributed transactions,'' \emph{Proc. ACM Manag. Data}, vol.~2, no.~1, mar 2024.

\bibitem{Bernstein:Book1987}
P.~A. Bernstein, V.~Hadzilacos, and N.~Goodman, \emph{Concurrency Control and Recovery in Database Systems}.\hskip 1em plus 0.5em minus 0.4em\relax Addison-Wesley, 1987.

\bibitem{SER:JACM1979}
C.~H. Papadimitriou, ``The serializability of concurrent database updates,'' \emph{J. ACM}, vol.~26, no.~4, pp. 631--653, 1979.

\bibitem{Lin:TOPLAS1990}
M.~P. Herlihy and J.~M. Wing, ``Linearizability: A correctness condition for concurrent objects,'' \emph{ACM Trans. Program. Lang. Syst.}, vol.~12, no.~3, pp. 463--492, 1990.

\bibitem{Framework:CONCUR2015}
A.~Cerone, G.~Bernardi, and A.~Gotsman, ``A framework for transactional consistency models with atomic visibility,'' in \emph{CONCUR'15}, vol.~42, 2015, pp. 58--71.

\bibitem{Fauna-ISO}
M.~F. Daniel~Abadi, ``Serializability vs ``strict'' serializability: The dirty secret of database isolation levels,'' \url{https://fauna.com/blog/serializability-vs-strict-serializability-the-dirty-secret-of-database-isolation-levels}, Accessed in April, 2024.

\bibitem{MongoDB-ISO}
{MongoDB}, \url{https://www.mongodb.com/docs/manual/core/transactions/}, Accessed in April, 2024.

\bibitem{PostgreSQL-ISO}
{PostgreSQL}, \url{https://www.postgresql.org/docs/current/transaction-iso.html}, Accessed in April, 2024.

\bibitem{MariaDB-ISO}
{MariaDB}, Accessed in April, 2024, \url{https://mariadb.com/kb/en/mariadb-transactions-and-isolation-levels-for-sql-server-users/}.

\bibitem{YugabyteDB-ISO}
{YugabyteDB}, \url{https://docs.yugabyte.com/preview/explore/transactions/isolation-levels/}, Accessed in April, 2024.

\bibitem{TiDB-ISO}
{TiDB}, \url{https://docs.pingcap.com/tidb/stable/transaction-isolation-levels}, Accessed in April, 2024.

\bibitem{Spanner:TOCS2013}
J.~C. Corbett, J.~Dean, M.~Epstein, A.~Fikes, C.~Frost, J.~J. Furman, S.~Ghemawat, A.~Gubarev, C.~Heiser, P.~Hochschild, W.~Hsieh, S.~Kanthak, E.~Kogan, H.~Li, A.~Lloyd, S.~Melnik, D.~Mwaura, D.~Nagle, S.~Quinlan, R.~Rao, L.~Rolig, Y.~Saito, M.~Szymaniak, C.~Taylor, R.~Wang, and D.~Woodford, ``Spanner: Google’s globally distributed database,'' \emph{ACM Trans. Comput. Syst.}, vol.~31, no.~3, p.~22, 2013.

\bibitem{MonkeyDB:OOPSLA2021}
R.~Biswas, D.~Kakwani, J.~Vedurada, C.~Enea, and A.~Lal, ``{MonkeyDB}: Effectively testing correctness under weak isolation levels,'' \emph{Proc. ACM Program. Lang.}, vol.~5, no. OOPSLA, 2021.

\bibitem{NCC:OSDI2023}
H.~Lu, S.~Mu, S.~Sen, and W.~Lloyd, ``{NCC}: Natural concurrency control for strictly serializable datastores by avoiding the timestamp-inversion pitfall,'' in \emph{OSDI'23}.\hskip 1em plus 0.5em minus 0.4em\relax USENIX Association, 2023, pp. 305--323.

\bibitem{AlgebraicLaw:CONCUR2017}
A.~Cerone, A.~Gotsman, and H.~Yang, ``Algebraic laws for weak consistency,'' in \emph{CONCUR'17}, vol.~85, 2017, pp. 26:1--26:18.

\bibitem{UnifiedTheory:JACM2004}
R.~C. Steinke and G.~J. Nutt, ``A unified theory of shared memory consistency,'' \emph{J. ACM}, vol.~51, no.~5, p. 800–849, 2004.

\bibitem{TSM:SIAM1997}
P.~B. Gibbons and E.~Korach, ``Testing shared memories,'' \emph{SIAM J. Comput.}, vol.~26, no.~4, pp. 1208--1244, aug 1997.

\bibitem{TAOMP:Book2012}
M.~Herlihy and N.~Shavit, \emph{The Art of Multiprocessor Programming, Revised Reprint}, 1st~ed.\hskip 1em plus 0.5em minus 0.4em\relax San Francisco, CA, USA: Morgan Kaufmann Publishers Inc., 2012.

\bibitem{Cassandra-LWT}
{Apache Cassandra}, ``Using lightweight transactions,'' \url{https://docs.datastax.com/en/cql-oss/3.3/cql/cql_using/useInsertLWT.html}, Accessed in April, 2024.

\bibitem{ScyllaDB-LWT}
ScyllaDB, ``Lightweight transactions,'' \url{https://opensource.docs.scylladb.com/stable/using-scylla/lwt.html}, Accessed in April, 2024.

\bibitem{PNUTS:VLDB2019}
B.~F. Cooper, P.~Narayan, R.~Ramakrishnan, U.~Srivastava, A.~Silberstein, P.~Bohannon, H.-A. Jacobsen, N.~Puz, D.~Weaver, and R.~Yerneni, ``{PNUTS} to sherpa: lessons from {Yahoo}!'s cloud database,'' \emph{Proceedings of the VLDB Endowment}, vol.~12, no.~12, pp. 2300--2307, 2019.

\bibitem{CosmosDB-LWT}
{Azure Cosmos DB}, ``{Azure Cosmos DB} for {Apache Cassandra} lightweight transactions with conditions,'' \url{https://learn.microsoft.com/en-us/azure/cosmos-db/cassandra/lightweight-transactions}, Accessed in April, 2024.

\bibitem{etcd-LWT}
etcd, ``etcd3 {API}: transaction,'' \url{https://etcd.io/docs/v3.4/learning/api/\#transaction}, Accessed in April, 2024.

\bibitem{MTC-Tool}
H.~Wei, J.~Xiao, N.~Yang, S.~Liu, Z.~Yifan, Y.~Chen, and A.~Pan, ``Artifacts for {MTC},'' \url{https://github.com/code-artifacts/mtc}, 2025.

\bibitem{MonoSAT:AAAI2015}
S.~Bayless, N.~Bayless, H.~H. Hoos, and A.~J. Hu, ``{SAT} modulo monotonic theories,'' in \emph{AAAI'15}, 2015, pp. 3702--3709.

\bibitem{horn2015faster}
A.~Horn and D.~Kroening, ``Faster linearizability checking via p-compositionality,'' in \emph{International Conference on Formal Techniques for Distributed Objects, Components, and Systems}, 2015, pp. 50--65.

\bibitem{mariadb-issue}
{Transactions on MariaDB-Galera violated Causal Consistency}, ``Issue \#609,'' \url{https://github.com/codership/galera/issues/609#issuecomment-1046763283}, Accessed in February, 2025.

\bibitem{MongoDB-Jepsen}
K.~Kingsbury, ``{Jepsen testing of MongoDB 4.2.6},'' \url{http://jepsen.io/analyses/mongodb-4.2.6}, Accessed in February, 2025.

\bibitem{Dgraph-Jepsen}
------, ``{Jepsen testing of Dgraph 1.1.1},'' \url{http://jepsen.io/analyses/dgraph-1.1.1}, Accessed in February, 2025.

\bibitem{pg-issue}
------, ``{Avoid update conflict out serialization anomalies},'' \url{https://git.postgresql.org/gitweb/?p=postgresql.git;a=commit;h=5940ffb221316ab73e6fdc780dfe9a07d4221ebb}, Accessed in February, 2025.

\bibitem{jepsen-cassandra}
------, ``Jepsen: Cassandra,'' \url{https://aphyr.com/posts/294-call-me-maybe-cassandra}, Accessed in April, 2024.

\bibitem{ConsMaude:TACAS2019}
S.~Liu, P.~C. {\"{O}}lveczky, M.~Zhang, Q.~Wang, and J.~Meseguer, ``Automatic analysis of consistency properties of distributed transaction systems in maude,'' in \emph{{TACAS} 2019}, ser. LNCS, vol. 11428, 2019, pp. 40--57.

\bibitem{Emme:EuroSys2024}
J.~Clark, A.~F. Donaldson, J.~Wickerson, and M.~Rigger, ``Validating database system isolation level implementations with version certificate recovery,'' in \emph{EuroSys'24}.\hskip 1em plus 0.5em minus 0.4em\relax ACM, 2024, p. 754–768.

\bibitem{VerIso}
S.~Ghasemirad, S.~Liu, C.~Sprenger, L.~Multazzu, and D.~Basin, ``{VerIso}: Verifiable isolation guarantees for database transactions,'' \emph{Proc. {VLDB} Endow.}, vol.~18, no.~5, pp. 1362 -- 1375, 2025.

\bibitem{jepsen}
K.~Kingsbury, ``Jepsen,'' \url{https://jepsen.io}, Accessed in February, 2024.

\bibitem{Knossos}
------, ``Knossos,'' \url{https://github.com/jepsen-io/knossos}, Accessed in December, 2023.

\bibitem{Metamorphic:arXiv2020}
\BIBentryALTinterwordspacing
T.~Y. Chen, S.~C. Cheung, and S.~Yiu, ``Metamorphic testing: {A} new approach for generating next test cases,'' \emph{CoRR}, vol. abs/2002.12543, 2020. [Online]. Available: \url{https://arxiv.org/abs/2002.12543}
\BIBentrySTDinterwordspacing

\bibitem{FaunaDB-Jepsen}
K.~Kingsbury, ``{Jepsen testing of FaunaDB 2.5.4},'' \url{https://jepsen.io/analyses/faunadb-2.5.4}, Accessed in February, 2025.

\bibitem{VoltDB-Jepsen}
------, ``{Jepsen testing of VoltDB 6.3},'' \url{https://jepsen.io/analyses/voltdb-6-3}, Accessed in February, 2025.

\bibitem{DBStorm:ISSTA2024}
K.~Li, S.~Weng, L.~Ni, C.~Yang, R.~Zhang, X.~Zhou, and A.~Zhou, ``{DBStorm}: Generating various effective workloads for testing isolation levels,'' in \emph{ISSTA 2024}.\hskip 1em plus 0.5em minus 0.4em\relax ACM, 2024, p. 755–767.

\bibitem{IsoVista}
L.~Gu, S.~Liu, T.~Xing, H.~Wei, Y.~Chen, and D.~A. Basin, ``{IsoVista}: Black-box checking database isolation guarantees,'' \emph{Proc. {VLDB} Endow.}, vol.~17, no.~12, pp. 4325--4328, 2024.

\end{thebibliography}


\clearpage
\appendices


\section{Correctness Proofs of the Verification Algorithms}
\label{section:appendix-proofs}

\begin{lemma} \label{lemma:no-fork-implies-total}
	If the graph $\G$ returned by \emph{\consdep}
	does not contain any $\WW$ loops or
	any instances of the \exception{} pattern,
	then it is a dependency graph of $\H$.
\end{lemma}

\begin{figure}[t]
	\centering
	\includegraphics[width = 0.30\textwidth]{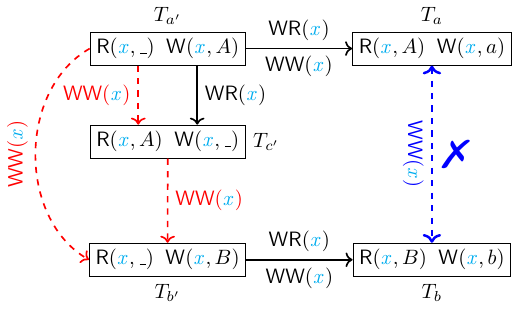}
	\caption{Transactions $T_{a}$ and $T_{c'}$ read the same value of the object $x$
		from the transaction $T_{a'}$ and then write different values.}
	\label{fig:pattern-existence}
\end{figure}

\begin{proof} \label{proof:no-fork-implies-total}
	Suppose that $\G$ does not contain any $\WW$ loops or
	any instances of the \exception{} pattern.
	We need to show that for each object $k \in \Key$,
	$\WW(k)$ is a strict total order on the set $\WriteTx_{x}$.
	Suppose by contradiction that
	there exists an object $x \in \Key$
	such that $\WW(x)$ is not a strict total order on $\WriteTx_{x}$.
	In the following, we identify an instance of the \exception{} pattern in $\G$.

	Since $\G$ does not contain any $\WW$ loops,
	$\WW(x)$ is a strict order.
	Therefore, $\WW(x)$ is not total.
	Consider two transactions $T_{a}$ and $T_{b}$
	such that they both write to $x$ with different values,
	but there are no $\WW(x)$ dependency edges between them in $\G$;
	see Figure~\ref{fig:pattern-existence}.
	If $T_{a}$ and $T_{b}$ read the value of $x$ from the same transaction,
	then we are done.
	Otherwise, consider the transactions $T_{a'}$ and $T_{b'}$
	($T_{a'} \neq T_{b'}$) from which
	$T_{a}$ and $T_{b}$ read the value of $x$, respectively.
	That is, $T_{a'} \rel{\WR(x)/\WW(x)} T_{a}$ and
	$T_{b'} \rel{\WR(x)/\WW(x)} T_{b}$
	(see line~\ref{line:ww-edges} of \consdep).
	It is clear that $T_{a'} \neq T_{b}$ and $T_{b'} \neq T_{a}$,
	since otherwise there would be a $\WW(x)$ dependency edge between $T_{a}$ and $T_{b}$.
	There are two cases regarding whether there is a $\WW(x)$ dependency edge
	between $T_{a'}$ and $T_{b'}$ in $\G$:
	\begin{itemize}
		\item Suppose there is a $\WW(x)$ dependency edge between $T_{a'}$ and $T_{b'}$ in $\G$,
			say $T_{a'} \rel{\WW(x)} T_{b'}$.
			Note that this $\WW(x)$ edge may be derived by transitivity
			(see line~\ref{line:ww-tc} of \consdep).
			Let $T_{c'}$ be a transaction \emph{directly following} $T_{a'}$
			in the $\WW(x)$ dependency chain from $T_{a'}$ to $T_{b'}$.
			That is, the edge $T_{a'} \rel{\WW(x)} T_{c'}$
			coincides with $T_{a'} \rel{\WR(x)} T_{b'}$
			(see line~\ref{line:ww-edges} of \consdep).
			As illustrated in Figure~\ref{fig:pattern-existence},
			we have $T_{a'} \rel{\WR(x)/\WW(x)} T_{c'} \rel{\WW(x)} T_{b'}$.
			It is possible that $T_{c'} = T_{b'}$, but we argue that
			\begin{itemize}
				\item $T_{c'} \neq T_{a}$.
				  Otherwise, we have $T_{a} = T_{c'} \rel{\WW(x)} T_{b'} \rel{\WW(x)} T_{b}$.
					Then by transitivity, $T_{a} \rel{\WW(x)} T_{b}$,
					contraditing the assumption that there is no $\WW(x)$ dependency edge
					between $T_{a}$ and $T_{b}$.
				\item $T_{c'} \neq T_{b}$.
				  Otherwise, we have $T_{b} = T_{c'} \rel{\WW(x)} T_{b'} \rel{\WW(x)} T_{b}$.
					Then by transitivity, $T_{b} \rel{\WW(x)} T_{b}$,
					contraditing the assumption that $\G$ contains no $\WW(x)$ loops.
			\end{itemize}

			Therefore, $T_{a'}$, $T_{c'}$, and $T_{a}$ forms an instance
			of the \exception{} pattern.
		\item If there is no a $\WW(x)$ dependency edge
			between $T_{a'}$ and $T_{b'}$ in $\G$,
			we apply the same argument above to $T_{a'}$ and $T_{b'}$.
	\end{itemize}
	Since the number of transactions in $\H$ is finite
	and $\H$ contains an initial transaction $\bot_{T}$
	which install initial values to all objects,
	we will eventually identify three transactions
	that form an instance of the \exception{} pattern.
\end{proof}


\begin{lemma} \label{lemma:acyclic-implies-total}
	If the graph $\G$ returned by \emph{\consdep} is acyclic,
	then it is a dependency graph of $\H$.
\end{lemma}

\begin{proof} \label{proof:fork}
	Suppose by contradiction that $\G$ is not a dependency graph of $\H$.
	By Lemma~\ref{lemma:no-fork-implies-total},
	$\G$ contains $\WW$ loops or instances of the \exception{} pattern.
	Since $\G$ is acyclic, there are no $\WW$ loops in $\G$.
	Thus, $\G$ must contain instances of the \exception{} pattern,
	as illustrated in Figure~\ref{fig:pattern}.
	However, in this case, there would be a cycle
	$T_{2} \rel{\RW(x)} T_{3} \rel{\RW(x)} T_{2}$ in $\G$.
\end{proof}

\begin{theorem}[Correctness of \checksser]
	\label{thm:correctness-checksser}
	\checksser{} returns \true{} if and only if $\H$ satisfies \sser.
\end{theorem}

\begin{proof} \label{proof:correctness-checksser}
	The proof proceeds in two directions.
	\begin{itemize}
		\item (``$\implies$'')
		  Suppose that \checksser{} returns \true.
			Then the graph $\G$ at line~\ref{line:checksser-call-consdep}
			of \checksser{} is acyclic.
		  By Lemma~\ref{lemma:acyclic-implies-total},
			$\G$ is a dependency graph of $\H$.
			By Definition~\ref{def:sser},
			$\H$ satisfies \sser.
		\item (``$\;\Longleftarrow\;$'')
		  Suppose that \checksser{} returns \false.
			Then the graph $\G$ at line~\ref{line:checksser-call-consdep}
			of \checksser{} is cyclic.
			Since $\G$ is a subgraph of any dependency graph of $\H$,
			no dependency graph of $\H$ is acyclic.
			By Definition~\ref{def:sser},
			$\H$ does not satisfy \sser.
	\end{itemize}
\end{proof}

Similarly, by Lemma~\ref{lemma:acyclic-implies-total} and Definition~\ref{def:ser},
we can establish the correctness of \checkser{}.
The proof is omitted here.

\begin{theorem}[Correctness of \checkser]
	\label{thm:correctness-checkser}
	\checkser{} returns \true{} if and only if $\H$ satisfies \ser.
\end{theorem}


Finally, we show the correctness of \checksi{}.
\begin{theorem}[Correctness of \checksi]
	\label{thm:correctness-checksi}
	\checksi{} returns \true{} if and only if $\H$ satisfies \si.
\end{theorem}


\begin{proof} \label{proof:si-theorem-mini}
	The proof proceeds in two directions.
	\begin{itemize}
		\item (``$\;\Longleftarrow\;$'')
		  We show that if \checksi{} returns \false,
			then $\H$ does not satisfy SI.
		  There are two cases:
			\begin{itemize}
				\item Suppose that \checksi{} returns \false{}
				at line~\ref{line:checksi-return-false} of \checksi.
				Then $\G$ contains an instance of the \exception{} pattern.
				By Lemma~\ref{lemma:fork-violates-si},
				$\H$ does not satisfy SI.
				\item Suppose that \checksi{} returns \false{}
				at line~\ref{line:checksi-acyclic} of \checksi.
				There are two cases regarding whether $\G$ contains a $\WW$ loop:
				\begin{itemize}
					\item If $\G'$ contains a $\WW$ loop,
					  then, by the construction of $\G'$
						(line~\ref{line:checksi-g-prime} of \checksi),
						the graph $\G$ at line~\ref{line:checksi-g-prime}
						also contains a $\WW$ loop.
						Since $\G$ is a {subgraph} of any dependency graph of $\H$,
						$\H$ does not admit any {legal} dependency graphs.
						By Definition~\ref{def:si},
						$\H$ does not satisfy SI.
					\item If $\G'$ contains no $\WW$ loops,
					  then, by the construction of $\G'$
						(line~\ref{line:checksi-g-prime} of \checksi),
						$\G$ does not contain any $\WW$ loop either.
						On the other hand, the graph $\G$ at line~\ref{line:checksi-g-prime}
						does not contain any instances of the \exception{} pattern.
						By Lemma~\ref{lemma:no-fork-implies-total},
						the graph $\G$ at line~\ref{line:checksi-g-prime} is a dependency graph of $\H$.
						Since $\G$ is a {subgraph} of any dependency graph of $\H$,
						$\G$ is the {unique} dependency graph of $\H$.
						Then by Definition~\ref{def:si},
						$\H$ does not satisfy SI.
				\end{itemize}
			\end{itemize}
		\item (``$\implies$'')
			If \checksi{} returns \true{}
			at line~\ref{line:checksi-acyclic} of \checksi,
			then $\G'$ at line~\ref{line:checksi-g-prime} of \checksi{} is acyclic.
			Therefore, $\G'$ contains no $\WW$ loops.
			By the line of reasoning above,
			$\G$ at line~\ref{line:checksi-g-prime} is a dependency graph of $\H$.
			Therefore, by Definition~\ref{def:si}, $\H$ satisfies SI.
	\end{itemize}
\end{proof}

\section{Correctness Proofs of the Optimized Verification Algorithms}
\label{section:appendix-opt-proof}




\begin{proof}[Proof of Lemma~\ref{lemma:widehat-rw}]
	\label{proof:widehat-rw}
	For the $S \rel{\widehat{\RW}} T'$ edge in $\G$,
	there exists a transaction $T$ such that
	$T \rel{\WR} S$ and $T \widehat{\rel{\WW}} T'$.
	Therefore, there exists a sequence of $\WW$ edges
	from $T$ to $T'$, denoted $T \rel{\WW} T_{1} \rel{\WW} T_{2}
	\rel{\WW} \cdots \rel{\WW} T'$ such that $S \rel{\RW} T_{1}$.
\end{proof}

The correctness of the optimized versions of \checksser{} and \checkser{}
directly follows from the following theorem.


\begin{proof}[Proof of Theorem~\ref{thm:G-Gbar-acyclic}]
	\label{proof:G-Gbar-acyclic}
	The proof proceeds in two directions.
	\begin{itemize}
		\item ``$\implies$'':
		  Suppose that $\G$ is acyclic.
		  Since $\edges_{\widehat{\G}} \subseteq \edges_{\G}$,
			$\widehat{\G}$ is also acyclic.
		\item ``\;$\Longleftarrow$\;'':
			Suppose by contradiction that $\G$ contains a cycle,
			denoted $\cycle$.
			We need to show that $\widehat{\G}$ also contains a cycle.

			If $\edges_{\cycle} \subseteq \edges_{\widehat{\G}'}$,
			then we are done.
			Otherwise, any edge in
			$\edges_{\cycle} \setminus \edges_{\widehat{\G}}$
			must be either a $\widehat{\WW}$ edge or a $\widehat{\RW}$ edge.
			We can transform $\cycle$ into a cycle within $\widehat{\G}$
			as follows (see Figure~\ref{fig:ww-tc}):
			\begin{itemize}
				\item For each $\widehat{\WW}$ edge in $\cycle$,
				  replace it with a sequence of $\WW$ edges
					from which the $\widehat{\WW}$ edge can be derived
					by transitivity.
				\item For each $S \rel{\widehat{\RW}} T'$ edge in $\cycle$,
				  replace it with a path $S \rel{\RW} T_{1} \rel{\WW} T_{2}
					\rel{\WW} \cdots \rel{\WW} T'$,
					as indicated by Lemma~\ref{lemma:widehat-rw}.
			\end{itemize}
	\end{itemize}
\end{proof}

Let $\widehat{\G}'$ be the induced graph of $\widehat{\G}$,
i.e., $\widehat{\G}' \gets (V_{\widehat{\G}},
  (\SO_{\widehat{\G}} \cup \WR_{\widehat{\G}} \cup \WW_{\widehat{G}})
	  \comp \RW_{\widehat{\G}}?)$.
The correctness of the optimized versions of \checksi{}
directly follows from the following theorem.


\begin{proof}[Proof of Theorem~\ref{thm:G'-Gbar'-acyclic}]
	\label{proof:G'-Gbar'-acyclic}
	The proof proceeds in two directions.
	\begin{itemize}
		\item ``$\implies$'':
		  Suppose that $\G'$ is acyclic.
		  Since $\edges_{\widehat{\G}'} \subseteq \edges_{\G'}$,
			$\widehat{\G}'$ is also acyclic.
		\item ``\;$\Longleftarrow$\;'':
			Suppose by contradiction that $\G'$ contains a cycle,
			denoted $\cycle'$.
			We need to show that $\widehat{\G}'$ also contains a cycle.

			If $\edges_{\cycle'} \subseteq \edges_{\widehat{\G}'}$,
			then we are done.
			Otherwise, we can transform $\cycle'$ into a cycle within $\widehat{\G}'$
			by replacing each edge in $\edges_{\cycle'} \setminus \edges_{\widehat{\G}'}$
			with edge(s) in $\widehat{\G}'$ as follows:
			\begin{enumerate}
				\item Consider a $\widehat{\WW}$ edge in $\cycle'$.
				  Replace it with a sequence of $\WW$ edges
					from which the $\widehat{\WW}$ edge can be derived
					by transitivity.
				\item Consider an $S' \rel{\SO \comp \widehat{\RW}} T'$ edge in $\cycle'$.
				  Suppose that it is derived from
					$S' \rel{\SO} S \rel{\widehat{\RW}} T'$.
				  As indicated by Lemma~\ref{lemma:widehat-rw},
					$S \rel{\widehat{\RW}} T'$ can be replaced by
					a path $S \rel{\RW} T_{1} \rel{\WW} T_{2} \rel{\WW} \cdots \rel{\WW} T'$,
					Therefore, $S' \rel{\SO \comp \widehat{\RW}} T'$
					can be replaced by the path
					$S' \rel{\SO \comp \RW} T_{1} \rel{\WW} T_{2} \rel{\WW} \cdots \rel{\WW} T'$
					in $\widehat{\G}'$.
				\item Consider a $\WR \comp \widehat{\RW}$ edge in $\cycle'$.
				  This is similar to case (2).
				\item Consider a $\WW \comp \widehat{\RW}$ edge in $\cycle'$.
				  This is similar to case (2).
				\item Consider a $S' \rel{\widehat{\WW} \comp \RW} T'$ edge in $\cycle'$.
				  Suppose that it is derived from $S' \rel{\widehat{\WW}} S \rel{\RW} T'$.
					The edge $S' \rel{\widehat{\WW}} S$
					can be replaced by a path $S' \rel{\WW} T_{1} \rel{\WW} T_{2}
					  \rel{\WW} \cdots \rel{\WW} T_{n} \rel{\WW} S$.
					Therefore, $S' \rel{\widehat{\WW} \comp \RW} T'$
					can be replaced by a path $S' \rel{\WW} T_{1} \rel{\WW} T_{2}
					  \rel{\WW} \cdots \rel{\WW} T_{n} \rel{\WW \comp \RW} T'$ in $\widehat{\G}'$.
				\item Consider a $S' \rel{\widehat{\WW} \comp \widehat{\RW}} T'$ edge in $\cycle'$.
				  Suppose that it is derived from $S' \rel{\WW} S \rel{\RW} T'$.
					Replace $S' \rel{\WW} S$ with a path $S' \rel{\WW} T_{1} \rel{\WW} T_{2}
					  \rel{\WW} \cdots \rel{\WW} T_{n} \rel{\WW} S$.
					Replace $S \rel{\RW} T'$ with a path $S \rel{\RW} T_{n+1} \rel{\WW} T_{n+2}
					  \rel{\WW} \cdots \rel{\WW} T_{n+m} \rel{\WW} T'$,
					as indicated by Lemma~\ref{lemma:widehat-rw}.
					Therefore, $S' \rel{\widehat{\WW} \comp \widehat{\RW}} T'$
					can be replaced by a path $S' \rel{\WW} T_{1} \rel{\WW} T_{2}
					  \rel{\WW} \cdots \rel{\WW} T_{n} \rel{\WW \comp \RW} T_{n+1}
					  \rel{\WW} T_{n+2} \rel{\WW} \cdots \rel{\WW} T_{n+m} \rel{\WW} T'$
					in $\widehat{\G}'$.
			\end{enumerate}
	\end{itemize}
\end{proof}

\section{Complexity Issues}
\label{section:appendix-complexity}

It is known that verifying
whether a given general history satisfies
\sser{}, \ser{} or \si{} is \npc~\cite{SER:JACM1979, Complexity:OOPSLA2019}.
In this section, we show that verifying
whether a given mini-transaction history
{\it without unique values}
satisfies \sser, \ser, or \si{} is also \npc.

\begin{theorem} \label{thm:sser-npc}
  The problem of verifying whether a given mini-transaction history
	without unique values
	satisfies \sser{} is \npc.
\end{theorem}

\begin{proof}
	Directly from \cite[Theorem 4.11]{TSM:SIAM1997}
	which shows that verifying whether an execution
	with read\&write operations only satisfies \lin{} is \npc.
\end{proof}

Our proof of the \nphardness{} of verifying \ser{}
for mini-transaction histories without unique values
relies on the \nphardness{} of verifying
\iso{Sequential Consistency} (\sccond). 
\sccond{} requires that all operations
appear to be executed in some sequential order
that is consistent with the session order.

\begin{definition}[Sequential Consistency]
	\label{def:sc}
	A history $\h$ is sequentially consistent if and only if
	there exists a permutation $\Pi$ of all the operations
	in $\h$ such that
	$\Pi$ preserves the program order of operations and
	follows the sequential semantics of each object.
\end{definition}

Note that when each transaction comprises only one operation,
\ser{} is the same as \sccond{}.

\begin{theorem} \label{thm:ser-npc}
  The problem of verifying whether a given mini-transaction history
	without unique values satisfies \ser{} is \npc.
\end{theorem}

\begin{proof}
	Directly from~\cite[Theorem 4.9]{TSM:SIAM1997}
	which shows that verifying whether an execution
	with read\&write operations only satisfies \sccond{} is \npc.
\end{proof}


The proof of the \nphardness{} of verifying \si{}
for mini-transaction histories without unique values
is much more involved.
It is heavily inspired by that
of~\cite[Theorem 3.2]{Complexity:OOPSLA2019}.\footnote{
	We largely follow the account of~\cite{Complexity:OOPSLA2019}
	and adapt it to our context when necessary.}
In the axiomatic framework of~\cite{Complexity:OOPSLA2019},
\si{} is characterized by two axioms: \prefixaxiom and \conflictaxiom.
The crucial difference is that in histories without unique values,
the write-read relation is not given as input.
In our context, a mini-transaction history (without unique values)
satisfies \si{} if there exist a write-read relation,
denoted by $\wrrelation$ to keep the notation consistent with~\cite{Complexity:OOPSLA2019},
and a commit order $\co$
(which is a strict total order on the set of transactions in $\H$)
such that the \prefixaxiom and \conflictaxiom axioms are satisfied.
In the following, we omit the $\RT$ relation in a history
since it is not relevant for \si.
An \emph{abstract execution} $\ae =(\T,\sorelation,\wrrelation,\co)$
is a history $(\T, \sorelation)$ associated with a write-read relation $\wrrelation$
and a commit order $\co$.
We use $R^{\ast}$ to denote the
reflexive and transitive closure of the relation $R$.

\begin{definition}[\prefixaxiom axiom~\cite{Complexity:OOPSLA2019}]
	An abstract execution $\mathcal{A}=(\T,\SO,\wrrelation,\co)$
	satisfies the \prefixaxiom axiom if and only if
	\begin{align*}
			&\forall x \in \Key.\; \forall t_1, t_2 \neq t_{1}, t_{3} \in \T. \\
					& \quad (\langle t_1,t_3\rangle \in \wrrelation(x)
					  \land t_2\vdash\writeevent(x,\_) \land
						\langle t_2,t_3\rangle\in\co^{\ast};(\wrrelation \cup \SO)) \\
					& \quad \implies \langle t_2,t_1\rangle\in\co. \\
	\end{align*}
\end{definition}

\begin{definition}[\conflictaxiom axiom~\cite{Complexity:OOPSLA2019}]
	An abstract execution $\ae =(\T, \SO, \wrrelation, \co)$
	satisfies the \conflictaxiom axiom if and only if
	\emph{\begin{align*}
				& \forall x, y \in \Key.\; \forall t_1, t_2, t_{3}, t_{4} \in \T.\; \\
						& \quad t_1\neq t_2\land\langle t_1,t_3\rangle\in\WR_x
						  \land t_2\vdash\writeevent(x,\_)\land t_3\vdash\writeevent(y,\_) \land \\
						& \quad t_4\vdash\writeevent(y,\_)\land\langle t_2,t_4\rangle\in\co^{\ast}
						  \land \langle t_4,t_3\rangle\in\co \\
						& \qquad \Rightarrow\langle t_2,t_1\rangle\in\co                                                                                           \\
		  \end{align*}}
\end{definition}

\begin{theorem} \label{thm:si-npc}
  The problem of verifying whether a given mini-transaction history
	without unique values
	satisfies \si{} is \npc.
\end{theorem}

\begin{proof}[Proof of Theorem~\ref{thm:si-npc}]
	\label{proof:si-npc}
  It is easy to see that the problem is in \np.
	To show \nphardness,
	we establish a reduction from boolean satisfiability.
	Let $\phi=D_1 \land \cdots \land D_m$
	be a CNF formula over boolean variables $x_1,\cdots,x_n$,
	where each $D_i$ is a disjunctive clause with $m_i$ literals.
	We use $\lambda_{ij}$ denote the $j$-th literal of $D_i$.
	We construct a mini-transaction history $h_{\phi}$ for $\phi$
	such that $\phi$ is satisfiable if and only if $h_{\phi}$ satisfies SI.

	The insight is to represent truth values of each variable and literal
	in $\phi$ with the polarity of the commit order
	between corresponding transaction pairs.
	Specifically, for each variable $x_k$,
	$h_{\phi}$ contains a pair of mini-transaction $a_k$ and $b_k$
	such that $x_k$ is false if and only if
	$\langle a_k,b_k\rangle\in \co$.
	For each literal $\lambda_{ij}$,
	$h_{\phi}$ contains a triple of mini-transactions
	$w_{ij},y_{ij}$, and $z_{ij}$ such that
	$\lambda_{ij}$ is false if and only if
	$\langle y_{ij}, z_{ij}\rangle \in \co$.

	The history $h_{\phi}$ should ensure that
	the $\co$ ordering corresponding to an assignment
	that makes the formula false form a cycle.
	To this end, we add all pairs
	$\langle z_{ij},y_{i,(j+1)\% m_i}\rangle$
	in the session order $\sorelation$.
	Consequently, an unsatisfied clause $D_i$
	leads to a cycle of the form
	$y_{i1}\xrightarrow{{\co}}z_{i1}\xrightarrow{{\sorelation}}y_{i2}
	\xrightarrow{{\co}}z_{i2}\cdots z_{im_i}\xrightarrow{{\sorelation}}y_{i1}$.


\begin{figure}
	\begin{subfigure}{.48\textwidth}
		\centering
		\includegraphics[width=.9\textwidth]{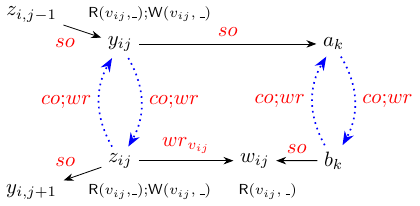}
		\caption{$\lambda_{ij}=x_k$}
		\label{fig:mini-si-npc-sub}
	\end{subfigure}\hfil
	\begin{subfigure}{.48\textwidth}
		\centering
		\includegraphics[width=.9\textwidth]{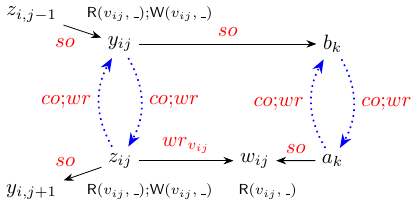}
		\caption{$\lambda_{ij}=\neg x_k$}
		\label{fig:mini-si-npc-sub-rev}
	\end{subfigure}
	\caption{Sub-histories in $h_{\phi}$ for literal $\lambda_{ij}$ and variable $x_k$}
	\label{fig:mini-si-npc-proof}
\end{figure}

	We use special sub-histories to ensure the consistency
	between the truth value of literals and variables.
	That is, $\lambda_{ij}=x_k$ is false if and only if $x_k$ is false.
	Figure~\ref{fig:mini-si-npc-sub} shows the sub-history
	associated to a positive literal $\lambda_{ij}=x_k$,
	while Figure~\ref{fig:mini-si-npc-sub-rev}
	shows the case of a negative literal $\lambda_{ij}=\neg x_k$.

	For a positive literal $\lambda_{ij}=x_k$ (Figure~\ref{fig:mini-si-npc-sub}),
	\begin{enumerate}
		\item we enrich the session order
			with the pairs $\langle y_{ij},a_k\rangle$ and $\langle b_k,w_{ij}\rangle$;
		\item we include reads and writes to a variable $v_{ij}$
		  in the transaction $y_{ij}$ and $z_{ij}$;
		\item we make $w_{ij}$ read $v_{ij}$; and
		\item we make all the read and written value the same.
	\end{enumerate}

	The steps (2)--(4) are specially designed for our reduction,
	different from that in the proof of~\cite{Complexity:OOPSLA2019}:
	\emph{it utilizes the fact that unique values are not required in such a history
	and ensures that every transaction in the sub-history
	is a mini-transaction.}

	\emph{We only need to determine the write-read relation
	between $y_{ij},z_{ij}$, and $w_{ij}$,
	since other transactions do not read or write the same variable.
	Specifically, we construct a write-read relation from $z_{ij}$ to $w_{ij}$.
	For transactions $y_{ij}$ and $z_{ij}$,
	which read and write the same variable with the same value,
	we require the direction of $\wrrelation$
	between $y_{ij}$ and $z_{ij}$ be
	the same as that of $\co$ between them.
	Otherwise, the history $h_{\phi}$ associated with
	these $\wrrelation$ and $\co$ relations
	cannot satisfy \prefixaxiom and \conflictaxiom.}

	\begin{lemma}
		\label{lemma:sub-histories}
		The special sub-histories enforce that
		if history $h_{\phi}$ satisfies SI,
		then there exist a write-read relation $\wrrelation$
		and a commit order $\co$ such that
		$\langle h_{\phi},\text{\wrrelation{}},\text{\co{}}\rangle$
		satisfies \prefixaxiom and \conflictaxiom and
		\begin{align*}
			& \langle a_k,b_k\rangle\in \co \text{ iff } \langle y_{ij},z_{ij}\rangle\in \co{} \text{ when } \lambda_{ij}=x_k, \text{and} \\
			& \langle a_k,b_k\rangle\in \co \text{ iff } \langle z_{ij},y_{ij}\rangle\in \co{} \text{ when } \lambda_{ij}=\neg x_k.
		\end{align*}
	\end{lemma}

	The proof of Lemma~\ref{lemma:sub-histories}
	and the remaining correctness proof of the reduction
	can be conducted similarly as those in~\cite{Complexity:OOPSLA2019}.
	We refer the interested reader to~\cite{Complexity:OOPSLA2019} for details.

\end{proof}

\section{End-to-End Checking Performance for SI}
\label{section:appendix-si-e2e}


\begin{figure*}[t]
	\centering
	\includegraphics[width = 0.95\textwidth]{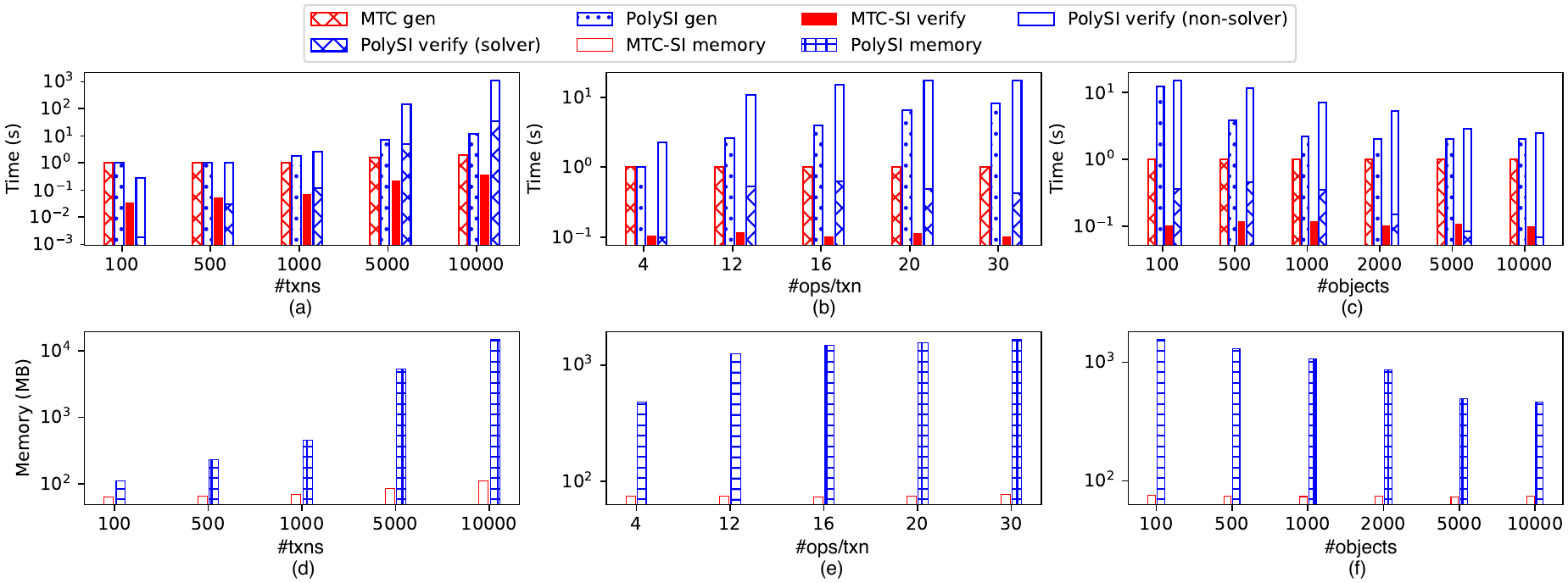}
	\caption{End-to-end checking performance,
	  with time decomposed into history generation and verification (on SI).}
	\label{fig:appendix-si-e2e}
\end{figure*}

\sitool with the MT workload generation
substantially outperforms PolySI with the GT workload generation
in terms of both time and memory
under varying concurrency levels,
as shown in Figure~\ref{fig:appendix-si-e2e}.

\section{Isolation Bugs Rediscovered by \ourtool}
\label{section:appendix-bugs}


\begin{figure*}[t]
	\centering
	\hfill
	\begin{subfigure}[b]{0.2\textwidth}
		\centering
		\includegraphics[height=\textwidth]{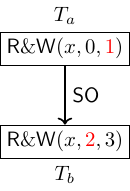}
		\caption{Cassandra: \textsc{read uncommitted} that violates SSER}
	\end{subfigure}
	\hfill
	\begin{subfigure}[b]{0.25\textwidth}
		\centering
		\includegraphics[width=\textwidth]{figs/bug-rep/stolon-bug.pdf}
		\caption{PostgreSQL: \textsc{write skew} that violates SER}
	\end{subfigure}
	\hfill
	\begin{subfigure}[b]{0.3\textwidth}
		\centering
		\includegraphics[width=\textwidth]{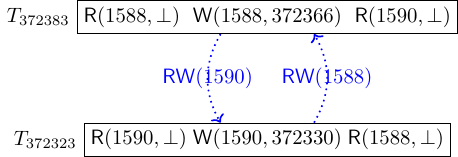}
		\caption{PostgreSQL:  \textsc{long fork} that violates SER}
	\end{subfigure}

	\begin{subfigure}[b]{0.2\textwidth}
		\centering
		\includegraphics[height=\textwidth]{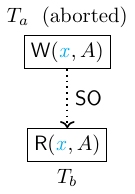}
		\caption{MongoDB: \textsc{read uncommitted} that violates SI}
	\end{subfigure}
	\hfill
	\begin{subfigure}[b]{0.3\textwidth}
		\centering
		\includegraphics[width=\textwidth]{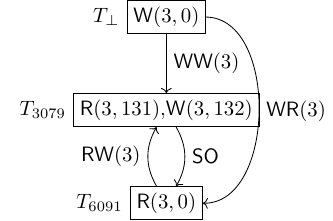}
		\caption{Dgraph: \textsc{causality violation} that violates SI}
	\end{subfigure}
	\hfill
	\begin{subfigure}[b]{0.25\textwidth}
		\centering
		\includegraphics[width=\textwidth]{figs/bug-rep/maria-mini.pdf}
		\caption{MariaDB Galera: \textsc{lost update} that violates SI}
	\end{subfigure}
	\hfill
	\caption{Rediscovered isolation bugs by \ourtool.}
	\label{fig:appendix-bugs}
\end{figure*}

Figure~\ref{fig:appendix-bugs} depicts
the isolation bugs detected by \ourtool
when checking six releases of five DBMSs
(see also Table~\ref{table:bugs}).

\end{document}